\documentclass[12pt,preprint]{aastex61}
\usepackage{color}

\usepackage{multirow}
\usepackage{graphicx}

\begin{document}
\title{Quasi-periodic oscillations in flares and coronal mass ejections associated with magnetic reconnection}
\author{Takuya Takahashi}
\affiliation{Kwasan and Hida Observatories, Kyoto University,Yamashina, Kyoto 607-8471, Japan.}
\email{takahasi@kusastro.kyoto-u.ac.jp}
\author{Jiong Qiu}
\affiliation{Department of Physics, Montana State University, Bozeman, MT 59717-3840, USA.}
\author{Kazunari Shibata}
\affiliation{Kwasan and Hida Observatories, Kyoto University,Yamashina, Kyoto 607-8471, Japan.}
%
%

\begin{abstract}
We propose a mechanism for quasi-periodic oscillations of both coronal mass ejections (CMEs) and flare loops as related to magnetic reconnection in eruptive solar flares. We perform two-dimensional numerical MHD simulations of magnetic flux rope eruption, with three different values of the global Lundquist number. In the low Lundquist number run, no oscillatory behavior is found. In the moderate Lunquist number run, on the other hand, quasi-periodic oscillations are excited both at the bottom of the flux rope and at the flare loop-top. In the high Lundquist number run, quasi-periodic oscillations are also excited; in the meanwhile, the dynamics become turbulent due to the formation of multiple plasmoids in the reconnection current sheet. In high and moderate Lundquist number runs, thin reconnection jet collide with the flux rope bottom or flare loop-top and dig them deeply. Steep oblique shocks are formed as termination shocks where reconnection jet is bent (rather than decelerated) in horizontal direction, resulting in supersonic back-flows. The structure becomes unstable, and quasi-periodic oscillation of supersonic back-flows appear at locally confined high-beta region at both the flux rope bottom and flare loop-top. We compare the observational characteristics of quasi-periodic oscillations in erupting flux ropes, post-CME current sheets, flare ribbons and light curves, with corresponding dynamical structures found in our simulation.
\end{abstract}

\keywords{Sun: flares --- Sun: coronal mass ejections --- solar-terrestrial relations --- solar wind --- Sun: heliosphere}

\section{Introduction}
Solar flares are among the largest explosions in the solar system where magnetic field energy of order of $10^{29}-10^{33}$ erg stored in active region corona is released through reconnection of magnetic field lines\citep{shibata2011}. Accompanying many solar flares, vast amount ($\sim 10^{13}-10^{17}$g) of coronal plasmas are ejected out into the interplanetary space with speeds of up to 3000 km s$^{-1}$ \citep{illing1986,aarnio2011,chen2011,webb2012}. Such plasma ejections are called coronal mass ejections or CMEs. CMEs disturb the plasma condition in the heliosphere and drive extreme space weather storms\citep{carr1859,tsuru1988,zhang2007,yer2012,gopa2014,haya2016,takahashi2016}. Many studies have been conducted to discuss the relationship between magnetic energy release during flares and acceleration of CMEs 
\citep{qiu2004,chen2010,karpen2012,lugaz2011}, as well as the propagation of CMEs in the heliosphere 
\citep{klein1982,cargill1996,manchester2004,qiu2007,vrsnak2013,liu2014,shiota2016,takahashi2017}.


During flares and CMEs, various intermittent and oscillatory features are observed and have been studied as important diagnostic tools for plasma properties and dynamics during the reconnection and eruption. Flare emission at the foot-points of reconnection-formed magnetic loops (flare loops) is intermittent in various wave bands such as hard X-ray (HXR), 
H$\alpha$, and white light. These signatures indicate the intermittent energy release process during flares \citep{nishizuka2009,nishizuka2010}. With the high spatio-temporal resolution of recent satellite observations, intermittent ejections of blob-like plasmas are ubiquitously found in solar flares. They are called ``plasmoids'' and thought to have helical magnetic field structure (magnetic flux ropes). Plasmoids are often observed in a linear bright structure beneath the CME flux rope, where the electric current is enhanced and magnetic reconnection occurs. Such a region of enhanced electric current is called the current sheet.

Recent numerical simulations show that plasmoids are generated within reconnection current sheet through successive progress of resistive tearing instability (plasmoid instability), when the Lundquist number of the current sheet exceeds a critical value of $S_c\simeq10^4$ \citep{tajima2002,loureiro2007}. When the plasmoid instability occurs in the current sheet, magnetic reconnection proceeds intermittently and is controlled by the dynamics of plasmoids\citep{shibata2001,janvier2011,loureiro2012,wang2015,huang2016}.
Two dimensional MHD simulation studies have revealed that the rate of magnetic reconnection is almost independent of the current sheet Lundquist number $S$ due to the nonlinear dynamics of the plasmoids once $S\gtrsim S_c$ is reached \citep{bhattacharjee2009}.

The observed quasi-periodic pulsations (QPPs) in HXR, EUV, and, optical emissions at the foot-points of flare loops indicate quasi-periodic precipitation of energetic electrons during flares. It has been proposed that QPPs are produced by the magnetohydrodynamic (MHD) modes trapped at the flaring loops, such as sausage mode and kink modes \citep{nakariakov2009}. MHD oscillation models of QPPs are applied to diagnose plasma properties of flare loops in solar active regions as well as in distant stars \citep{nakariakov2004}.
Another mechanism has also been proposed to explain the QPPs. Using high-resolution MHD simulations, \citet{takasao2016} have shown that flare loops can support localized non-linear oscillations at the loop top driven by the collision of reconnection jets, which are fast confined down-flows accelerated by magnetic reconnection. 
Recently, \citet{brannon2015} reported quasi-periodic oscillations of flare ribbons in a sawtooth-like pattern observed by the
{\em Interface Region Imaging Spectrograph}(IRIS). They discussed Kelvin-Helmholtz or Tearing mode instability in the coronal current sheet as the cause of the oscillatory pattern of the flare ribbons.

Quasi-periodic oscillations of the speed of CME flux ropes are also reported in coronagraph observations with LASCO on board SOHO \citep{krall2001}. \citet{shanmugaraju2010} reported that the leading edge of CMEs oscillates with a period of $\sim100$ minutes, which increases with time as the CME propagates. \citet{michalek2016} studied the statistical distribution of the CME oscillation and found their average speeds and periods to be $87$ km s$^{-1}$ and $241$ minutes, respectively. \citet{michalek2016} discussed that the model of the global oscillation of flux ropes proposed by \citet{cargill1994} gives a reasonable explanation if a thin flux tube geometry is assumed. \citet{lee2015} studied radial and azimuthal oscillation of nine halo CMEs observed by LASCO C3. They report that the instantaneous radial velocity varies quasi-periodically with period ranging from 24 to 48 minutes, and that the oscillations in seven events are associated with distinct azimuthal wave modes with wave number m=1 (asymmetric oscillation). They discussed that such a rapid oscillation of distant CMEs cannot be explained by traditional views of global MHD oscillations of flux ropes, and proposed that another nonlinear oscillation might be responsible, such as periodic shedding of Alfvenic vortices. Recently, \citet{li2016} reported the quasi-periodic oscillation of the post-CME current sheet observed with SDO/AIA. They reported the oscillation period of $\sim 11$ min and discussed that the oscillation is consistent with a fast propagating MHD kink wave. 
They also reported that the oscillation continued for longer than two hours.

In this paper, we propose a mechanism for quasi-periodic oscillations of flare loops and CMEs. We have found, in our high-resolution MHD simulations, local nonlinear oscillations driven by the collision of reconnection outflows with the CME flux rope as well as the flare loops. In section 2, we describe the setting of the MHD simulation. In section 3, we present the global time evolution of our simulation. In section 4, we study the asymmetric oscillation behind the flux rope caused by the collision of reconnection jet. In section 5, we discuss the physical quantities that govern the oscillation period. In section 6, we study the oscillatory behavior of the CME radial and expansion speeds. In section 7, we compare an observed quasi-periodic pulsations in an M-class flare with the oscillation reproduced in the simulation.

\section{The numerical method}
We conduct 2D numerical MHD simulations to study the dynamics of flux rope eruption driven by magnetic reconnection at the current sheet beneath the rope based on the catastrophe model \citep{forbes1990,lin2000,mei2012}. The pre-eruption field is composed of three parts, a current carrying flux rope in the corona, its image current 
below the photosphere, and the back ground potential arcade by a quadrupole. When the magnetic quadrupole strength is smaller than a critical value, 
there is no equilibrium solution for this magnetic system, resulting in an eruption. We start our simulation with the magnetic quadrupole 
slightly weaker than the critical strength, and track the dynamics involved in the ensueing flux rope eruption and magnetic reconnection at the current sheet beneath.

\subsection{The system equations}
We numerically solved 2D resistive MHD equations as shown below.
\begin{equation}
\frac{\partial\rho}{\partial t}+\nabla\cdot\left(\rho{\bf v}\right)=0
\end{equation}
\begin{equation}
\frac{\partial}{\partial t}(\rho{\bf v})+\nabla\cdot\biggl[\rho{\bf v}{\bf v}+\bigl(p+\frac{B^2}{8\pi}\bigr){\bf I}-\frac{{\bf B}{\bf B}}{4\pi}\biggr]-\rho{\bf g}=0
\end{equation}
\begin{equation}
\frac{\partial {\bf B}}{\partial t}+c\nabla\times{\bf E}=0
\end{equation}
\begin{equation}
\frac{\partial}{\partial t}\bigl(\frac{p}{\gamma-1}+\frac{1}{2}\rho{\bf v}^2+\frac{{\bf B}^2}{8\pi}\bigr)+\nabla\cdot\bigl[\bigl(\frac{\gamma}{\gamma-1}p+\frac{1}{2}\rho{\bf v}^2\bigr){\bf v}+\frac{c}{4\pi}{\bf E}\times{\bf B}\bigr]=0
\end{equation}
\begin{equation}
{\bf E}=\eta_0{\bf J}-\frac{1}{c}{\bf v}\times{\bf B}
\end{equation}
\begin{equation}
\frac{\partial \phi}{\partial t}+c_h^2\nabla\cdot{\bf B}+\frac{c_h^2}{c_p^2}\phi=0
\end{equation}
and
\begin{equation}
{\bf J}=\frac{c}{4\pi}\nabla\times{\bf B}
\end{equation}
with $\rho,~p,~\bf{B}$ and $\bf{v}$ being mass density, gas pressure, magnetic field and velocity, respectively. ${\bf g}$ is the gravitational acceleration in the solar atmosphere. $\bf I$ is the unit tensor, and ${\bf J}$ and ${\bf E}$ are electric current density and electric field, respectively. $c$ is the speed of light in vacuum. $\eta_0$ is uniform magnetic diffusivity and $\gamma=1.2$ is the ratio of specific heats. 

An additional quantity $\phi$ is introduced in the magnetic induction equation to remove numerical $\nabla\cdot{\bf B}$ based on the method proposed by \citet{dedner2002}. $c_h$ and $c_p$ are parameters which control the speed of advection and diffusion of numerical $\nabla\cdot{\bf B}$ whose values are adjusted so that numerical $\nabla\cdot{\bf B}$ does not ruin the physics. The numerical scheme is Harten-Lax-van Leer-Discontinuities (HLLD) approximate Riemann solver (which is a shock capturing scheme) \citep{miyoshi2005} with second-order total variation diminishing (TVD) Monotonic Upstream-Centered Scheme for Conservation Laws (MUSCL) and second order time integration. We did three simulations with different values of magnetic diffusivity $\eta_0$. We call them Run A, B and C, respectively. The Lundquist number $S=V_{FR}h/\eta_0$ for runs A, B and C are $S_A=2.8\times10^3$, $S_B=5.5\times10^3$ and $S_C=2.8\times10^4$, respectively, where $h$ and $V_{FR}$ are the initial height of the flux rope and the Alfven speed at the center of the flux rope, respectively. The values of $S$ and $M_q$ (the strength of the magnetic quadrupole, see the next section) used in the three different simulation runs are shown in Table 1.

\subsection{The initial magnetic field structure}
The magnetic field in the XZ plane can be expressed in terms of a magnetic flux function (the $y$-component of the magnetic vector potential) 
$\Psi$ as $B_x=-\frac{\partial\Psi}{\partial z}$ and $B_z=\frac{\partial\Psi}{\partial x}$.
\begin{figure}
\includegraphics[width=1.0\textwidth]{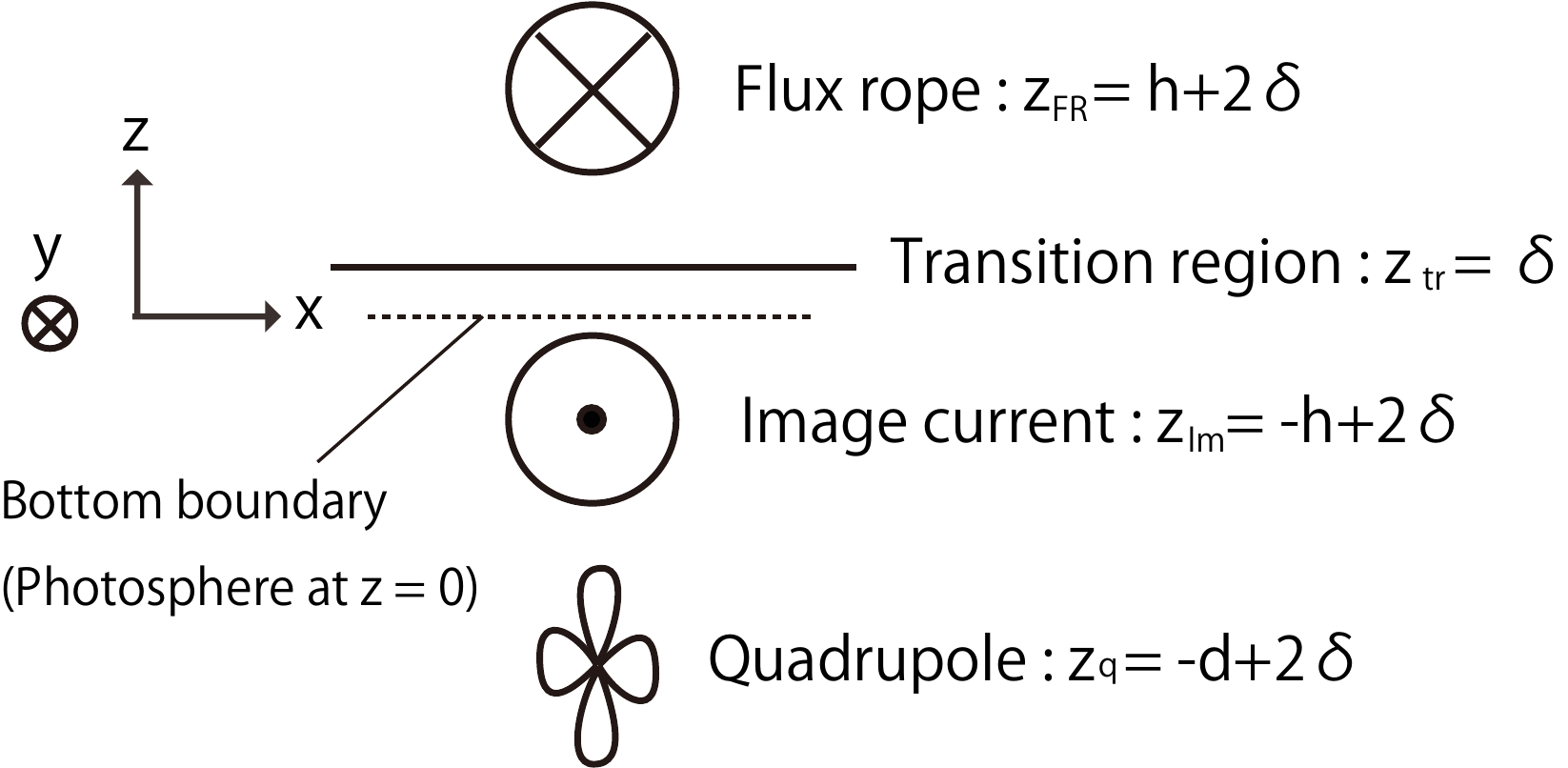}
\caption{The schematic figure of the Initial magnetic field structure. The flux rope, its image current and the buried quadrupole are shown. The transition region between the corona and the chromosphere is at a height of $z=\delta$. The bottom boundary of the simulation box is at $z=0$.}
\label{flare}
\end{figure}
The initial magnetic field is constructed with two oppositely directed electric current, the current carried in the flux rope in the corona and its 
image current below the photosphere, and a magnetic quadrupole. Figure 1 schematically shows the way initial magnetic structure is constructed. The magnetic flux function $\Psi$ can be expressed 
as the sum of the above three contributions as follows,
\begin{equation}
\Psi(x,z)=\Psi_{FR}(x,z)+\Psi_{Im}(x,z)+\Psi_{q}(x,z),
\end{equation}
where $\Psi_{FR}$, $\Psi_{Im}$ and $\Psi_q$ are magnetic flux functions by the flux rope current in the corona, its image current beneath the photosphere, and the magnetic quadrupole, respectively. 
The dense chromosphere is placed below the transition region at the height of $z_{tr}=\delta$ to prevent numerical instability at the bottom. The flux rope center is located at $z_{FR}=h+2\delta$. The image current is located at $z_{Im}=-h+2\delta$, and the magnetic quadrupole is buried at $z_q=-d+2\delta$ below the photosphere. 

The $y$-component of the electric current density within the flux rope $J_{FR,y}$ is a function of the radial distance from the flux rope center 
$r=\sqrt{x^2+(z-z_{FR})^2}$ as follows,
\begin{equation}
J_{FR,y}(r)=\left\{ \begin{array}{ll}
\frac{2I_0}{\pi{}R_0^2}\bigl(1-(\frac{r}{R_0})^2\bigr)~~(r < R_0) \\
0  ~~~~~~~~~~~~(r > R_0),
\end{array} \right.
\end{equation}
where $I_0$ is the total electric current flowing in the flux rope and $R_0$ is the radius of the flux rope. 
The Ampere's circuital law around the magnetic flux rope is written as $2\pi R_0B_0 = \frac{4\pi}{c}I_0$, with $B_0$ being the magnetic field strength at the 
edge of the flux rope.
The magnetic flux function $\Psi_{FR}$ is then written as follows,
\begin{equation}
\Psi_{FR}(r)=\left\{ \begin{array}{ll}
-B_0R_0\bigl[(\frac{r}{R_0})^2-\frac{1}{4}(\frac{r}{R_0})^4\bigr]~~(r < R_0) \\
B_0R_0\bigl[\frac{3}{4}-\log\frac{r}{R_0}\bigr]  ~~~~~~~~~~~(r > R_0)  .
\end{array} \right.
\end{equation}

The azimuthal component of the flux rope magnetic field $B_{FR,\phi}$ at the distance of $r$ from the flux rope center is 
\begin{equation}
B_{FR,\phi}(r)=-\frac{d\Psi_{FR}}{d r}=B_0\frac{r}{R_0}\big(2-\frac{r^2}{R_0^2}\big) .
\end{equation}

Then, we add the axial component of magnetic field within the flux rope $B_y(r)$ so that the flux rope be force-free in itself. 
The force-free condition is written as follows,
\begin{equation}
J_{FR,y}(r)B_{FR,\phi}(r)-J_{FR,\phi}(r)B_{FR,y}(r)=0,
\end{equation}
where $J_{FR,\phi}$ is the azimuthal component of the electric current density in the flux rope. $J_{FR,\phi}$ is the 
gradient of the newly added axial magnetic field $B_y$,
\begin{equation}
J_{FR,\phi}(r)=\frac{c}{4\pi}\frac{dB_{FR,y}}{dr}. 
\end{equation}

Substituting Equations (9), (11), (13) and $I_0=\frac{1}{2}cR_0B_0$ into Equation (12), the required axial magnetic field is obtained as follows,
\begin{equation}
B_{FR,y}(x,z)=\left\{ \begin{array}{ll}
B_0\sqrt{\frac{2}{3}\bigl[5-12(\frac{r}{R_0})^2+9(\frac{r}{R_0})^4-2(\frac{r}{R_0})^6\bigr]}~~(r < R_0) \\
0 ~~~~~~~~~~~(r > R_0)
\end{array} \right.
\end{equation}

We note that a uniform distribution of $J_y$ inside the flux rope is used in previous studies \citep{nishida2013}. We found that in 
such a case, the region around the boundary between the current carrying flux rope and outer coronal potential arcade suffers 
non-negligible numerical instability, which eventually destroys the calculation especially in the case of a low plasma beta. 
In this study, we modified the $J_{FR,y}$ distribution to be continuous so that we can avoid those numerical instabilities.
\citet{titov2014} used the same parabolic current distribution, although in a torus, not a straight cylinder.

The total amount of the image current is $I_0$ directed to negative $y$-direction and is buried at the depth of $z_{Im}=-h+2\delta$ beneath the photosphere. The magnetic flux function $\Psi_{Im}$ is a function of $r=\sqrt{x^2+(z-z_{Im})^2}$ as follows,
\begin{equation}
\Psi_{Im}(r)=-B_0R_0\log\frac{r}{R_0}.
\end{equation}

Finally, the flux function of the magnetic quadrupole of strength $m_q$ buried at the depth of $z=z_q=-d+2\delta$ is as follows,
\begin{equation}
\Psi_{q}(x,z)=-\frac{4\pi}{c}\frac{m_q}{4\pi r^2}\frac{x^2-(z-z_q)^2}{r^2},
\end{equation}
with $r=\sqrt{x^2+(z-z_q)^2}$.
We define a non-dimensional magnetic quadrupole strength $M_q$ by $m_q=I_0d^2M_q$, and rewrite Equation (16) as follows,
\begin{equation}
\Psi_{q}(x,z)=-\frac{B_0R_0d^2M_q}{2r^2}\frac{x^2-(z-z_q)^2}{r^2}.
\end{equation}

%

When $\delta$ is negligible compared with $h$ and $d$, the equilibrium height of the current carrying flux rope is obtained by eliminating the net Lorentz force on the flux rope current $I_0$, 
\begin{equation}
I_0\times\bigl(B_{Im,x}(0,h)+B_{q,x}(0,h)\bigr)=0,
\end{equation}
where $B_{Im,x}$ and $B_{q,x}$ are x component of the magnetic field originated from the image current and the quadrupole, respectively.

From Equations (15) and (17), $B_{Im,x}(0,h)$ and $B_{q,x}(0,h)$ are expressed as follows,
\begin{equation}
B_{Im,x}(0,h)=-\frac{\partial\Psi_{Im}(0,h)}{\partial h}=-B_0\frac{R_0}{2h},
\end{equation}
\begin{equation}
B_{q,x}(0,h)=-\frac{\partial\Psi_q(0,h)}{\partial h}=B_0\frac{R_0d^2M_q}{(h+d)^3}.
\end{equation}

Substituting Equations (19) and (20) into Equation (18), we get the equilibrium condition in a form of a cubic equation of $\zeta=h/d$ as follows,
\begin{equation}
\bigl(1+\zeta\bigr)^3-2M_q\zeta=0.
\end{equation}
The positive solution ($\zeta>0$) of Equation (21) gives the equilibrium height of the flux rope with respect to a given strength of the
magnetic quadrupole $M_q$. When $M_q$ is greater than the critical value $M_c=27/8$, the cubic Equation (21) has two positive roots $\zeta_1$ and $\zeta_2$ 
($\zeta_1<\zeta_2$). $\zeta_1$ and $\zeta_2$ give stable and unstable equilibrium height of the flux rope, respectively. When $M_q=M_c$, the two 
solutions degenerate at $\zeta_c=1/2$. If $M_q$ is smaller than $M_c$, there is no positive root in Equation (21), so no equilibrium can be found in this case. Similar discussions in the case of a dipole background field is given in \citet{forbes1990}.

$h=\frac{d}{2}$ ($\zeta=\zeta_c$) is used in all the simulation Runs. The quadrupole strength in all the simulation runs are set to be slightly smaller than the critical value ($M_q=0.95M_c$ for Run C and $M_q=0.92M_c$ for Run A and B) so that the flux rope is gradually accelerated upward right after the simulation starts. The Lundquist number $S=V_{FR}h/\eta_0$ and quadrupole strength $M_q$ for each Run is summarized in Table 1.

\begin{table}
\begin{center}
\caption{Simulation parameters for Runs A, B and C \label{tbl-1}}
\begin{tabular}{crrrrrrrrrrr}
\hline
\hline
\multicolumn{1}{c}{$~~~~$} & \multicolumn{1}{c}{$S$} &\multicolumn{1}{c}{$M_q$}\\
\hline
Run A &$~~~$  $2.8\times10^3$ &$~~~$  $0.92M_c$ &$~~~$ \\
Run B &$~~~$  $5.5\times10^3$ &$~~~$  $0.95M_c$ &$~~~$ \\
Run C &$~~~$  $2.8\times10^4$ &$~~~$  $0.92M_c$ &$~~~$ \\
\hline
\end{tabular}
\end{center}
\end{table}

\subsection{The gravitationally stratified atmosphere and non-uniform mesh}
The simulated atmosphere is composed of two layers, the high temperature corona with temperature $T_{cor}=1.5\times10^6$K and the
low temperature chromosphere with temperature $T_{chr}=1.5\times10^4$ K. The transition region between the corona and the chromosphere is at $z=\delta=7.0\times10^8$ cm with thickness $w_{tr}=2.1\times10^8$ cm. The explicit formula of the temperature distribution is as follows,
\begin{equation}
T(z)=T_{chr}+\frac{1}{2}(T_{cor}-T_{chr})\bigl(1+\tanh\frac{z-\delta}{w_{tr}}\bigr)
\end{equation}

The atmosphere is stratified under the solar gravitational acceleration,
\begin{equation}
g_z(z)=\left\{ \begin{array}{ll}
-g_0\frac{R_{\odot}^2}{(z+R_{\odot})^2}~~(0.5\delta<z<2R_{\odot})\\
0 ~~~~~~~~~~~(z<0.5\delta, z>2R_{\odot}),
\end{array} \right.
\end{equation}
with $R_{\odot}=7.0\times10^{10}$ cm and $g_0=2.74\times10^4$ cm s$^{-2}$ being the solar radius and the gravitational acceleration at the photosphere, respectively. The $x$ and $y$ components of gravitational acceleration are set to be 0. 
We calculate the initial distribution of electron number density $n_e(z)$ by the equation of hydrostatic equilibrium as follows,
\begin{equation}
n_e(z)m_Hg_z(z)-\frac{d}{dz}\bigl(2k_Bn_e(z)T(z)\bigr)=0,
\end{equation}
where $m_H$ and $k_B$ are the hydrogen mass and the Boltzmann constant, respectively. The electron number density at the base of the corona is set to be $3\times10^{10}$ cm$^{-3}$ in our simulation. The plasma $\beta$ (the ratio of gas pressure to magnetic pressure) is $~0.01$ around the flux rope. 
Figure 2 shows the electron number density of the initial atmosphere in log scale with magnetic field lines.

\begin{figure}
\includegraphics{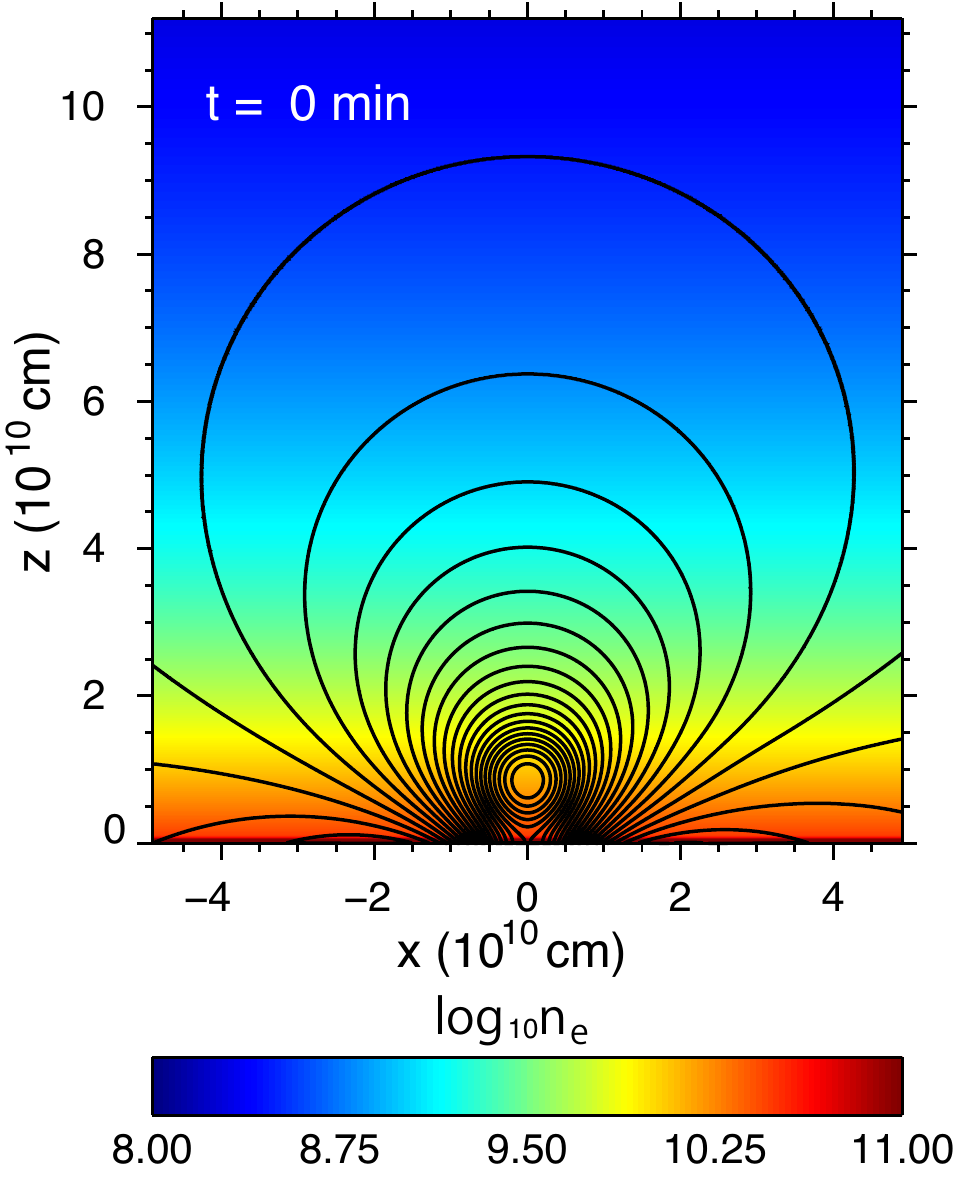}\centering
\caption{The initial magnetic field structure (solid black contours) and electron number density of gravitationally stratified atmosphere in log scale (color shadings) for Runs B and C.}
\label{vp}
\end{figure}

The simulation box is $x \in [-3R_{\odot} , 3R_{\odot}]$ and $z \in [0, 6R_{\odot}]$ discretized with non-uniformly arranged $N_x\times N_z=800\times1600$ grid points. The finest grid size in $x$-direction in Run A, B and C are set to be $\delta x = 6.7\times10^7$ cm, $3.3\times10^7$ cm and $8.3\times10^6$ cm, respectively, so that we can resolve the thinest current sheet beneath the erupting flux rope with $\sim$10 grid points. The current sheet is resolved with uniformly arranged 80 grid points near the $x=0$ plane. The box below the height of $0.6R_{\odot}=4.2\times10^{10}$ cm is discretized with uniformly arranged 1200 grid points with the grid size
of $\delta z=3.5\times 10^7$ cm. We use sparse grids in the outer space so that we can neglect unwanted numerical effects on the flux rope eruption dynamics from outer simulation boundaries. The distribution of the grid size ($\delta x$ and $\delta z$) at a given location in Run B is shown in Figure 3 (a) and (b). We applied reflective boundary condition in $x=\pm 3R_{\odot}$ and $z=6_{\odot}$ planes, and we fixed physical quantities at the bottom boundary. 

\begin{figure}
\includegraphics[width=1.0\textwidth]{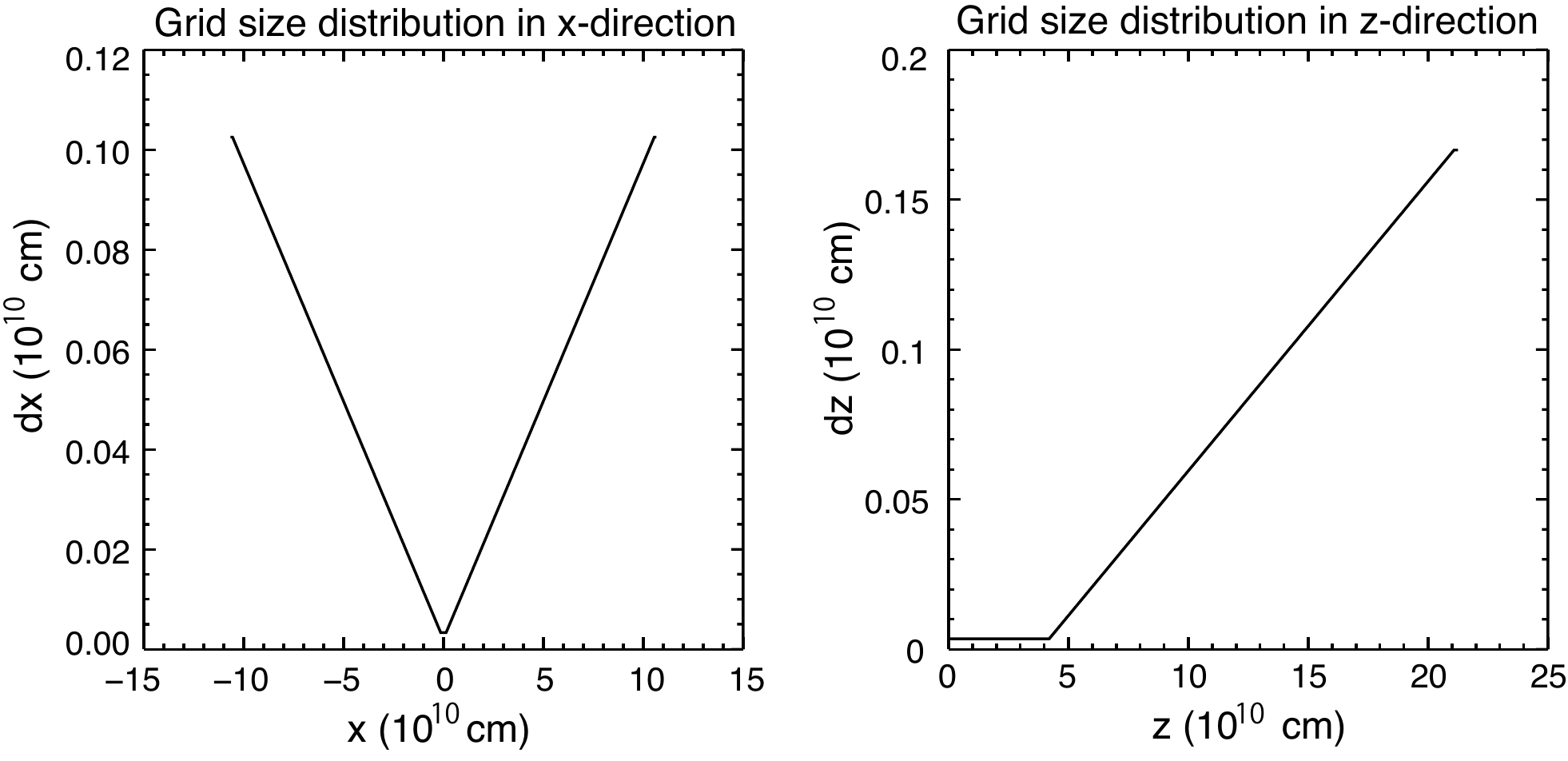}
\caption{Distribution of numerical grid size in x and z directions in Run B.}
\label{flare}
\end{figure}

\section{The numerical results in Runs A, B and C}
In this section, we compare the global time evolution of flux rope eruption in numerical simulation Runs A, B and C. Basically, the difference between the three runs are the Lundquist number of the flux rope (or in other words, magnetic diffusivity in the simulation). Simulation parameters are  summarized in Table 1).
Figure 4 shows the time evolution of electron number density distribution in log scale in Runs A, B and C. In the first row, we can see a globally propagating shock front launched from the erupting flux rope site in all the Runs. The second and third rows are the time when the flux rope is at the height 
of $\sim2.5\times10^{10}$ cm and $\sim6\times10^{10}$ cm, respectively in Runs A, B and C. The global electron number density and magnetic field structure at $t=95$ minutes in Run A is very similar to those at $t=138$ minutes in Run B and at $t=133$ minutes in Run C shown in the bottom panels of Figure 4. High density upward plasma jet in the current sheet (reconnection jet) collide with the erupting flux rope, and forms high density envelope of CME, while the downward reconnection jet collide with a closed magnetic loop at the bottom forming the flare loop. At the top of the CME, reconnection jet plasmas that have propagated upward along the sides of the CME flux rope collide and form a region with high density plasma. We also see some fine structures at the bottom and sides of the flux ropes in Runs B and C.

\begin{figure}
\includegraphics[width=1.0\textwidth]{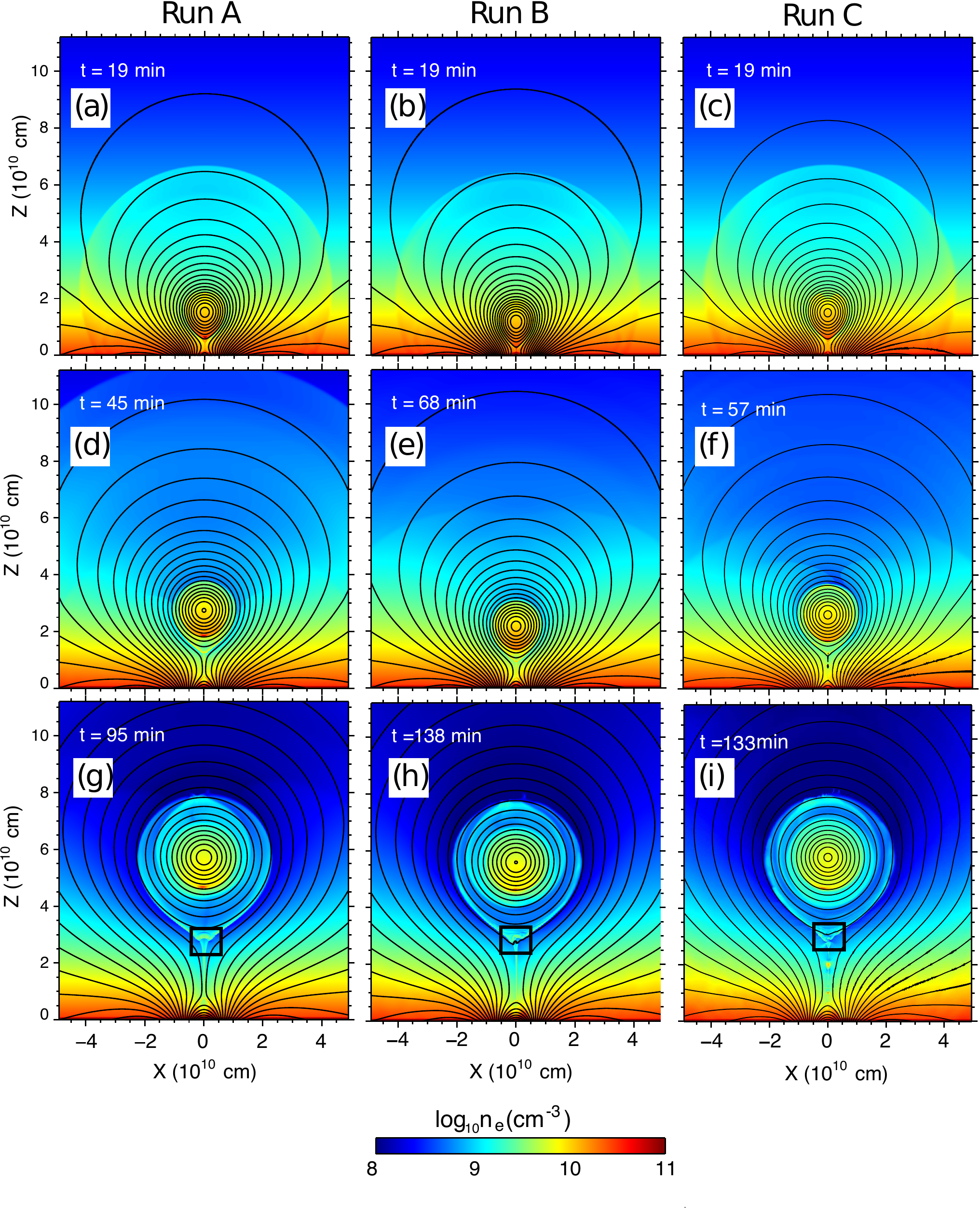}
\caption{Time evolution of the electron number density in log scale in Runs A, B and C. The first row shows the snap shot at $t=19$ minutes in all the Runs. The second and third rows are the time when the flux rope is at the height of $\sim2.5\times10^{10}$ cm and $\sim6\times10^{10}$ cm, respectively in Runs A, B and C. The major difference between the three simulation runs are the global Lundquist number $S$ as shown in Table 1.}
\label{flare}
\end{figure}

\begin{figure}
\includegraphics[width=1.0\textwidth]{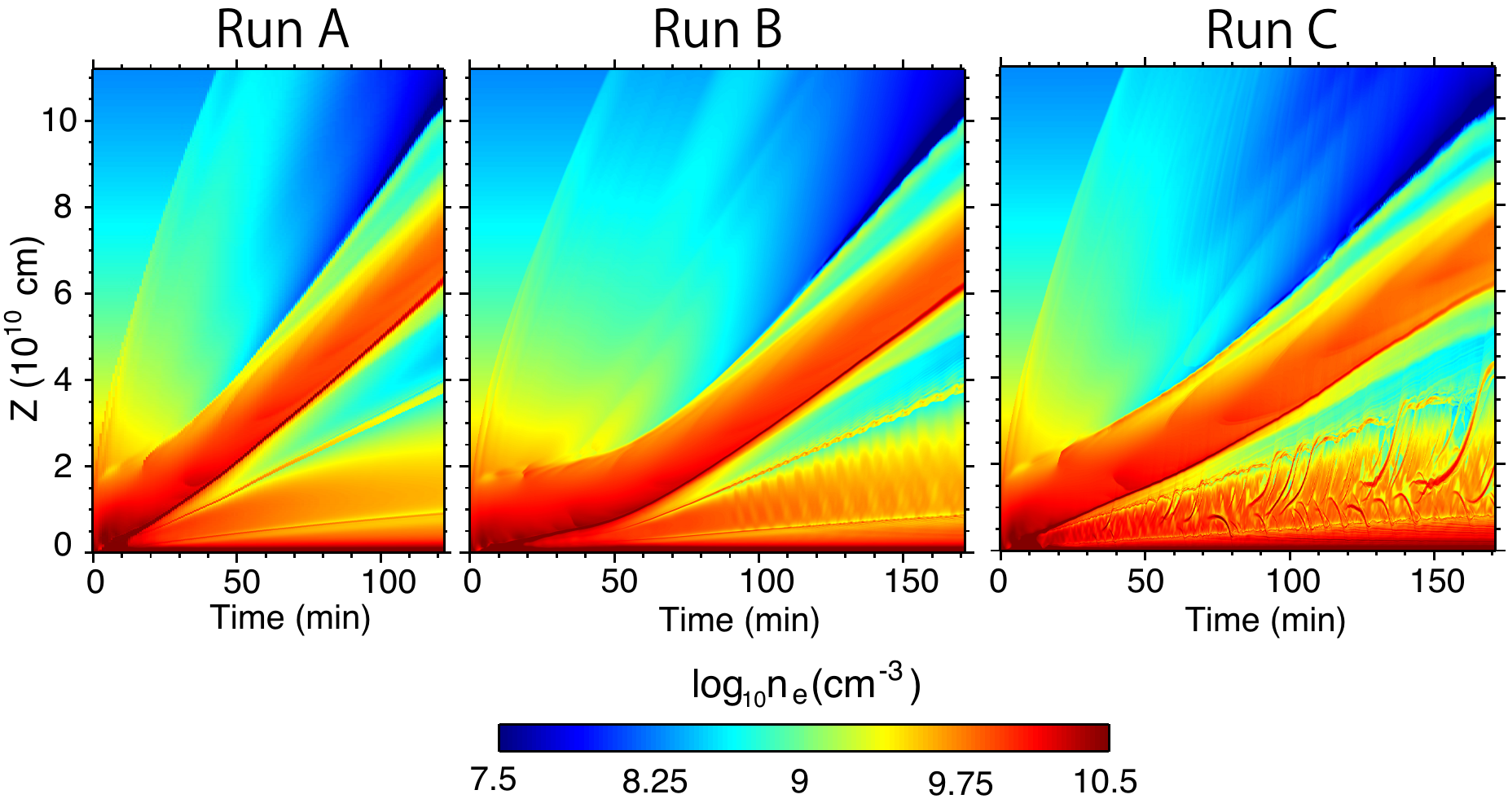}
\caption{Time-distance diagram of electron number density in log scale along the line $x=0$ in Runs A, B and C. The time range of the plot is selected so that the flux rope in each Run reaches the height of $\sim 1R_{\odot}=7\times10^{10}$ cm. We can see that global density structure in Run A evolves much faster than in Runs B and C. Also, we can see oscillating density structures around the current sheet beneath the flux rope in Runs B and C, but not in Run A. Plasmoids appear at $t\sim70$ minutes after the start of the simulation in Run C. The major difference between the three simulation runs are the global Lundquist number $S$ as shown in Table A.1.}
\label{flare}
\end{figure}

In Figure 5, we show the time-distance plots of electron number density $n_e$ in log scale along $x=0$ line in Runs A, B and C. The upper-most disturbance in each panels propagating upward is the MHD fast mode shock wave train launched from the erupting flux rope site. The flux rope with relatively high plasma density is being accelerated upward. Figure 6 schematically shows the correspondence between dynamical structures in 2.5D MHD simulation and those seen in Figure 5. The reconnection jet plasmas start to surround the current carrying flux rope right after the magnetic reconnection starts. At times $t\simeq50$ minutes in Run A and $t\simeq80$ minutes in Runs B and C, the reconnection jet plasma reaches the top of the flux rope and begin to accumulate above the flux rope. The boundary between the corona and the accumulated plasma forms the new leading edge of the CME. The coronal plasma over the leading edge is being evacuated. The reconnected plasma below the flux rope and the current sheet is separated by a thin layer of hot and dense plasma shocked by the termination shock. The lower end of the current sheet is attached to the flare loops with the termination shock in between. In Run A, the plasma structure around the reconnection current sheet seems to be laminar, while in Run B, we see some oscillations in density structure near the termination shocks and in the current sheet beginning at time $t\sim 80$ minutes. In Run C, in addition to the oscillations appeared at time $t\simeq 25$ minutes, high density plasma blobs start to appear at time $t\simeq 40$ minutes and being ejected out from the current sheet. They are ``plasmoids''. As time goes on, many plasmoids appear within the sheet and sometimes collide with each other before they are ejected out from the current sheet. Such dynamical process of plasmoids within reconnection current sheet has been discussed to be important for the understanding of magnetic reconnection speed and intermittency in previous studies. In this study, on the other hand, we focus more on the the dynamics of oscillations than those of plasmoids.

\begin{figure}
\includegraphics[width=1.0\textwidth]{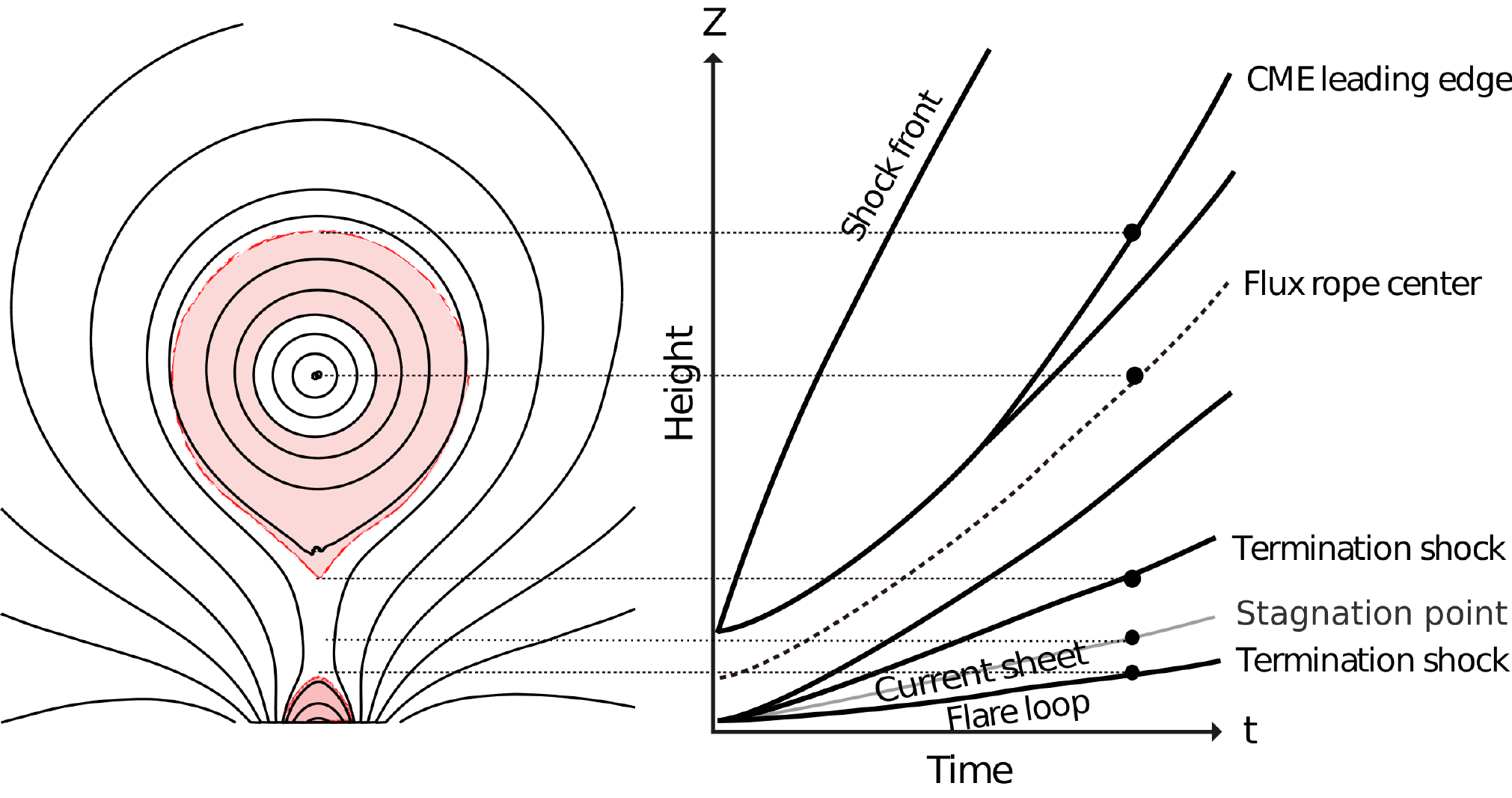}
\caption{Correspondence of dynamical structures in 2.5D MHD simulation and those seen in time-distance plots along the line $x=0$. The coronal MHD shock propagate outward soon after the eruption starts (the uppermost curve in the right panel). The reconnection jet plasma propagated along the sides of CME flux rope reaches the top of the CME and forms the CME leading edge. Termination shocks are seen at both sides of the reconnection current sheet. In between the two termination shocks, we also show the location of the stagnation point with a translucent curve, where vertical flow speed vanishes. Under the lower termination shock, flare loops are formed.}
\label{flare}
\end{figure}

\begin{figure}
\includegraphics[width=1.0\textwidth]{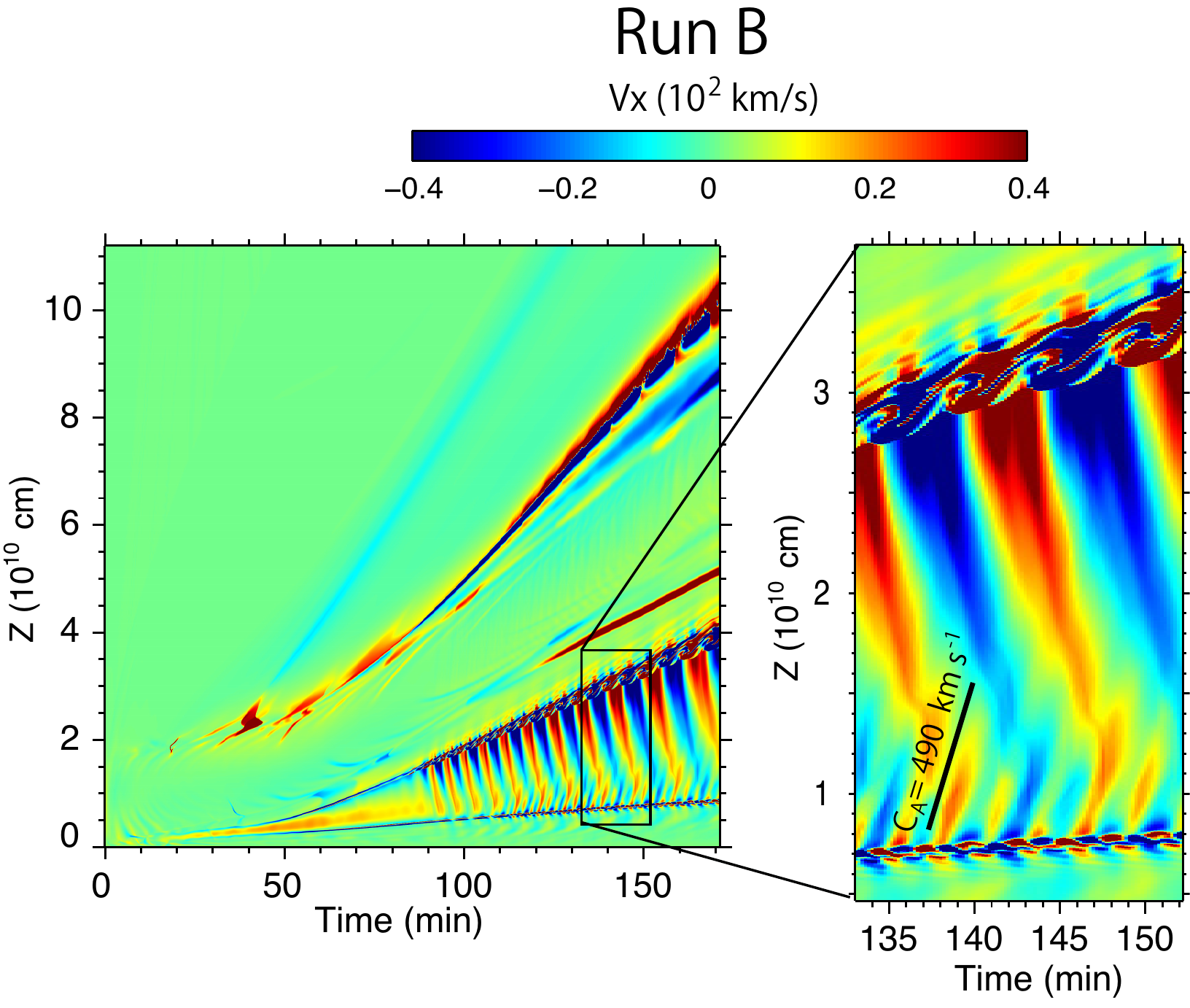}
\caption{Time-distance diagram of $v_x$ along the $x=0$ line in Run B. The close up look of the plot near the current sheet during the time between $t=133$ minutes and $t=152$ minutes are also shown. The phase speed of Alfven wave in the inflow region around the stagnation point at time $t=138$ minutes was 490 km s$^{-1}$ and is shown as a solid line in the right panel.}
\label{flare}
\end{figure}

\begin{figure}
\includegraphics[width=0.7\textwidth]{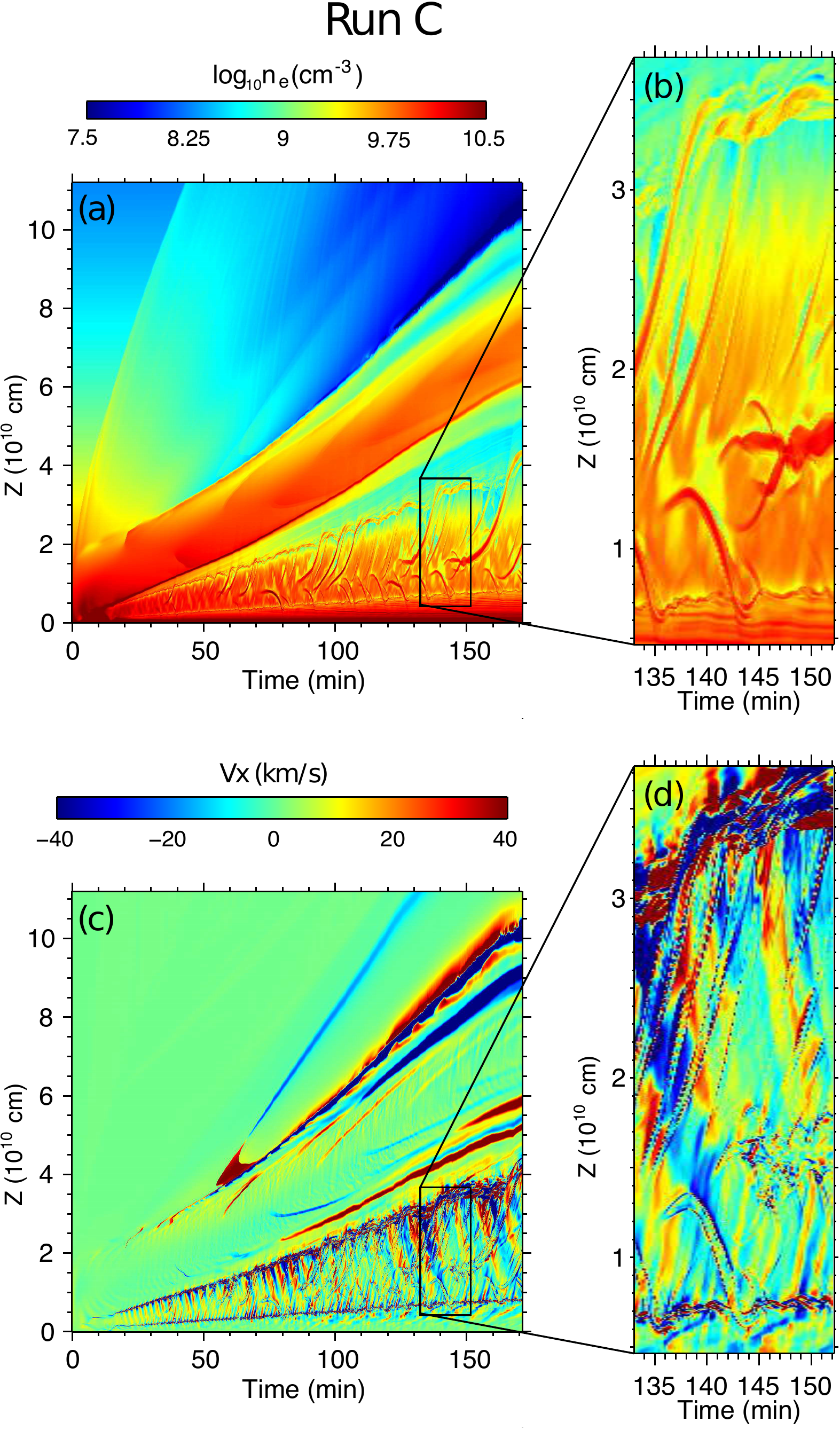}
\caption{Time-distance diagram of $\rho$ (panel (a)) and $v_x$ (panel (c)) along the $x=0$ line in Run C. The close up looks of the plot near the current sheet during the time between $t=133$ minutes and $t=152$ minutes are also shown as panels (b) and (d), respectively. In panel (b), dynamical behavior (formation, collision and ejection) of plasmoids in the current sheet can be seen clearly. In panel (d), the oscillations at both ends of the current sheets can be seen although the current sheet is plasmoid unstable.}
\label{flare}
\end{figure}

The oscillation is seen more clearly in the time-distance plot of $v_x$ along $x=0$ line in Run B as shown in Figure 7. We notice that not only the upper part of the termination shock (or current sheet) but also the lower part of the termination shock (flare loop top) oscillates. The oscillation at the upper and lower part of the current sheet begins at time $t\simeq90$ minutes and $t\simeq115$ minutes, respectively. The oscillatory pattern at the upper end of the current sheet travels downward towards the flare loops. The oscillatory pattern at the lower end of the current sheet on the other hand propagate upward. The phase speed of the Alfven wave in the inflow region (out of the current sheet) at the height of stagnation point (the point along the current sheet where $v_z$ vanishes) was 490 km s$^{-1}$ at time $t=138$ minutes, which is similar to the propagation speed of upward traveling wave pattern. This shows that the current sheet oscillation is originated from both the flux rope bottom and the flare loop top where collimated reconnection jets collide with closed magnetic loops. The leading edge of the CME also oscillate horizontally after time $t\simeq110$ minutes. This corresponds to the arrival of oscillating reconnection jet plasma that have traveled along the sides of the flux rope. 
Figure 8 shows the time-distance diagram of $\rho$ (panel (a)) and $v_x$ (panel (c)) along the $x=0$ line in Run C. The oscillatory behavior can be seen both around the current sheet and at the CME leading edge, although the structure of the current sheet in Run C becomes more complex due to the plasmoid instability (see also panels (b) and (d) for the close ups around the current sheet during the time between $t=133$ minutes and $t=152$ minutes).

\section{Oscillation at the bottom of the flux rope}
\begin{figure}
\includegraphics[width=1.0\textwidth]{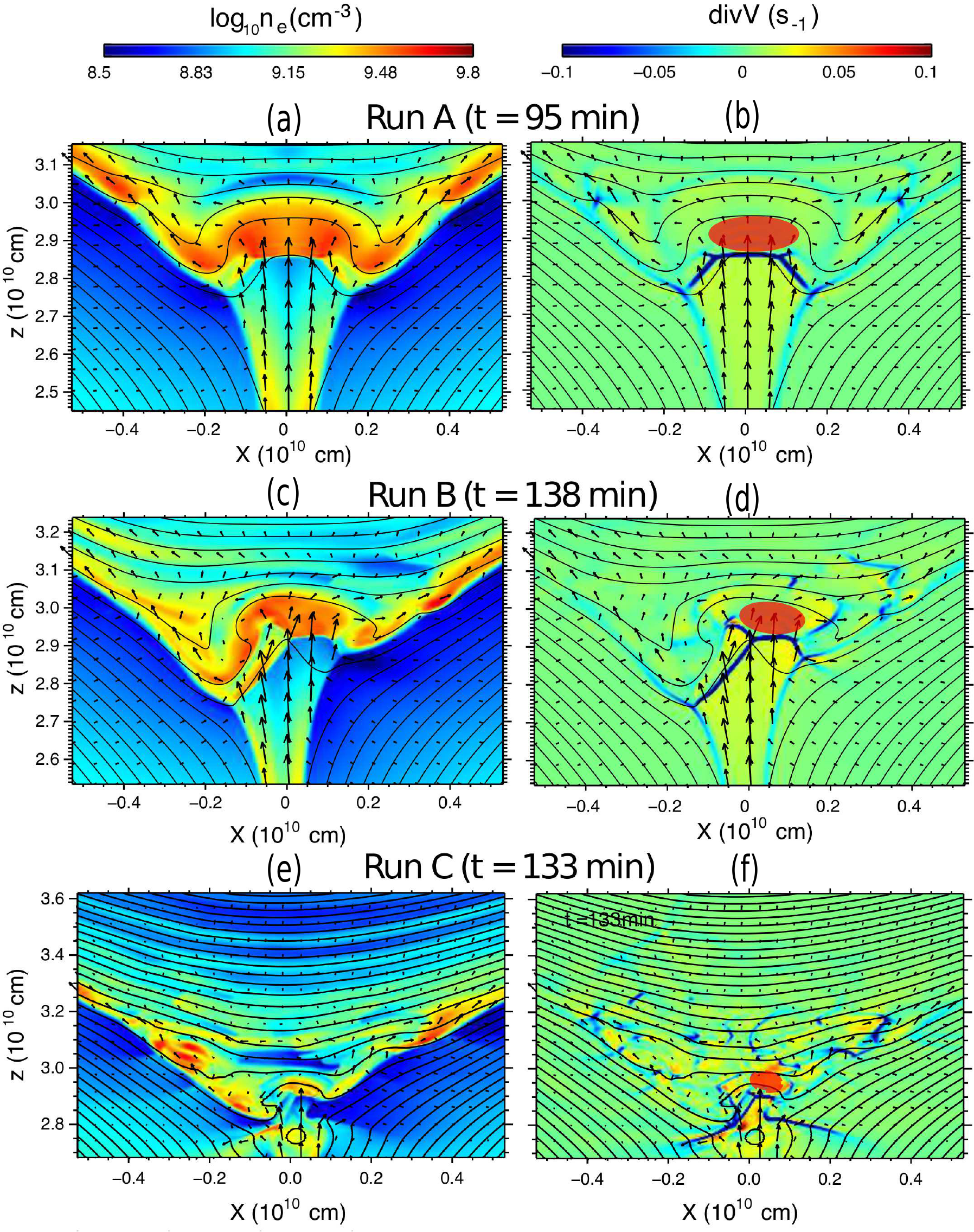}
\caption{The electron number density ($n_e$) and the divergence of the velocity $\nabla\cdot{\bf v}$ at the buffer region at time $t=95$ minutes, $138$ minutes and $133$ minutes after the start of the simulation in Runs A, B and C, respectively. The black arrows denotes the plasma flow field and the black contours denote magnetic field lines. The locations of the buffer region for each case are shown as red translucent ellipses in the panels in the right row.}
\label{flare}
\end{figure}

In this section, we focus on the dynamics of the oscillation at the upper end of the current
sheet (or the bottom of the flux rope) discussed in the previous section, by comparing the 
three simulation Runs. Although the lower end of the current sheet (at the top of flare loops) also oscillates due 
to the collision of reconnection jet, the oscillation at the bottom of the flux rope has a larger spatial scale,
so that fine structures involved in the oscillations can be numerically well-resolved. Figure 9 shows the 
electron number density in logarithmic scale ($\log n_e$) and the divergence of the velocity $\nabla\cdot{\bf v}$ 
at the bottom of the flux rope at times $t=95$ minutes, $t=138$ minutes and $t=133$ minutes in Runs A, B and C, respectively. 
The field of views of each panel in Figure 9 are indicated by white rectangles in Figure 4 (g), (h) and (i), respectively. For brevity,
in the following text, we call the region where upward reconnection jets collide with the bottom of the flux rope the ``buffer region'', 
and call the oscillation the ``buffer oscillation''.

In Run A, the termination shock is composed of two oblique shocks and one horizontal shock in between. These are observed as regions of 
concentration of the negative $\nabla\cdot {\bf v}$ (blue linear region) in Figure 9 (b). In Run A, the termination shock
consisting of three standing shocks is symmetric about the $z$-axis, and this symmetric structure is stable throughout the simulation and does not
exhibit oscillation signatures. In Run B, before the time $t\simeq 90$ minutes, the structure of the termination shock at the buffer region 
is similar to that in Run A, also consisting of two oblique shocks and one horizontal shock with a spatial symmetry. But after $t\simeq90$ minutes, 
the termination shock becomes asymmetric and suddenly begins to oscillate (the buffer oscillation). After the start of
the buffer oscillation, the termination shock is no longer symmetric, and oscillation continues throughout the rest of the simulation. 
Figure 9 (e) and (f) show $\log n_e$ and $\nabla\cdot {\bf v}$ at time $t=133$ minutes in Run C. The shock structure in this
run is more complicated due to the combination of the buffer oscillation and the collision of plasmoids ejected out from the current 
sheet in an intermittent manner. In the following section, we focus on the dynamics of the buffer oscillation in Run B.

\begin{figure}
\includegraphics[width=1.0\textwidth]{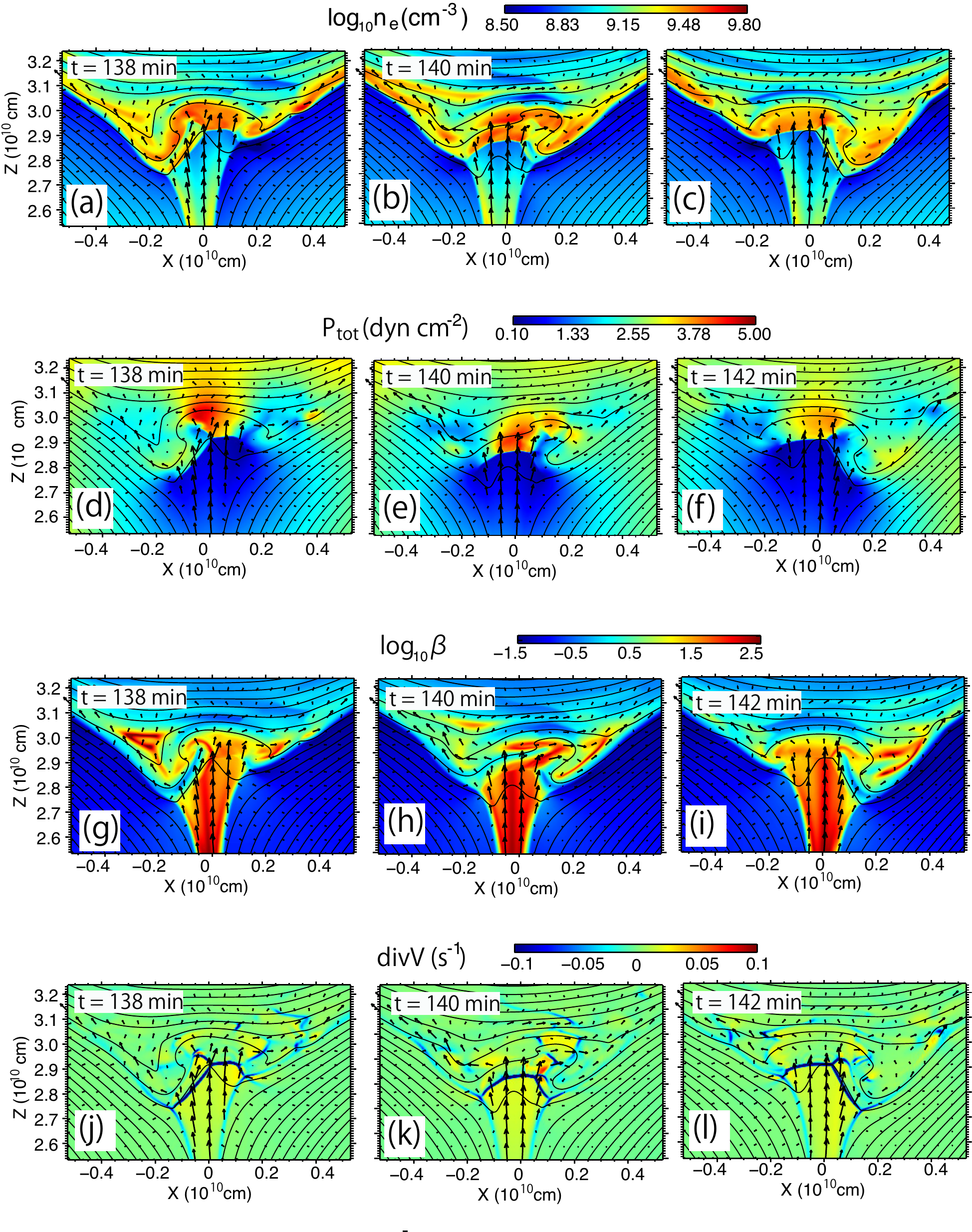}
\caption{Plasma quantities ($\log n_e$, $p+\frac{1}{8\pi}{\bf B}^2$, $\beta$ and $\nabla\cdot {\bf v}$ in color shadings) during the half cycle of the buffer oscillation in Run B. The magnetic field lines are shown as black contours and the velocity field is shown with black arrows.}
\label{flare}
\end{figure}

Figure 10 shows various plasma quantities at the buffer region during the half cycle of the oscillation
in Run B. Namely, electron number density in log scale ($\log n_e$), total pressure ($P_{tot}$), 
plasma beta ($\beta$) and the divergence of velocity ($\nabla\cdot {\bf v}$) are shown in each row.
At time $t=138$ minutes, the structure of the termination shock becomes asymmetric, and a long oblique shock is formed on 
the left (negative $x$) side of the buffer region (Figure 10 (j)). The reconnection jet enters the long oblique shock and collides with 
a long linear region of strong magnetic pressure and low plasma $\beta$, and is diverted to the right without significant deceleration 
(Figure 10 (g)). The diverted flow experiences the shock once again before it joins together with the reconnection jet that passes 
through the horizontal termination shock (Figure 10 (a) and (j)). The merged plasma is then accelerated mainly to the positive x direction 
by gas pressure gradient (Figure 10 (a) and (d)). Then, at time $t=140$ minutes, the 
low $\beta$ region (and the associated oblique shock) is stretched further towards the positive $x$ direction; consequently, the reconnection 
jet that enters the buffer is also turned to the positive $x$ direction (Figure 10 (e)). As a result, plasmas are being accumulated on 
the right (positive $x$) side of the buffer region, entering the next half-cycle of the oscillation.

\section{Evolution of the buffer oscillation period during the flux rope eruption}
In this section, we quantitatively discuss the evolution of the buffer oscillation and its relation with the
progress of the flux rope eruption. For this purpose, we have developed a method to automatically detect 
the height of the CME leading edge ($z_{LE}$), the center of the flux rope ($z_{FR}$), the upper and lower 
termination shocks ($z_{TS+}$ and $z_{TS-}$), and the stagnation points of the reconnection outflow ($z_{SP}$) 
along the $x=0$ line. 

\begin{figure}
\includegraphics[width=1.0\textwidth]{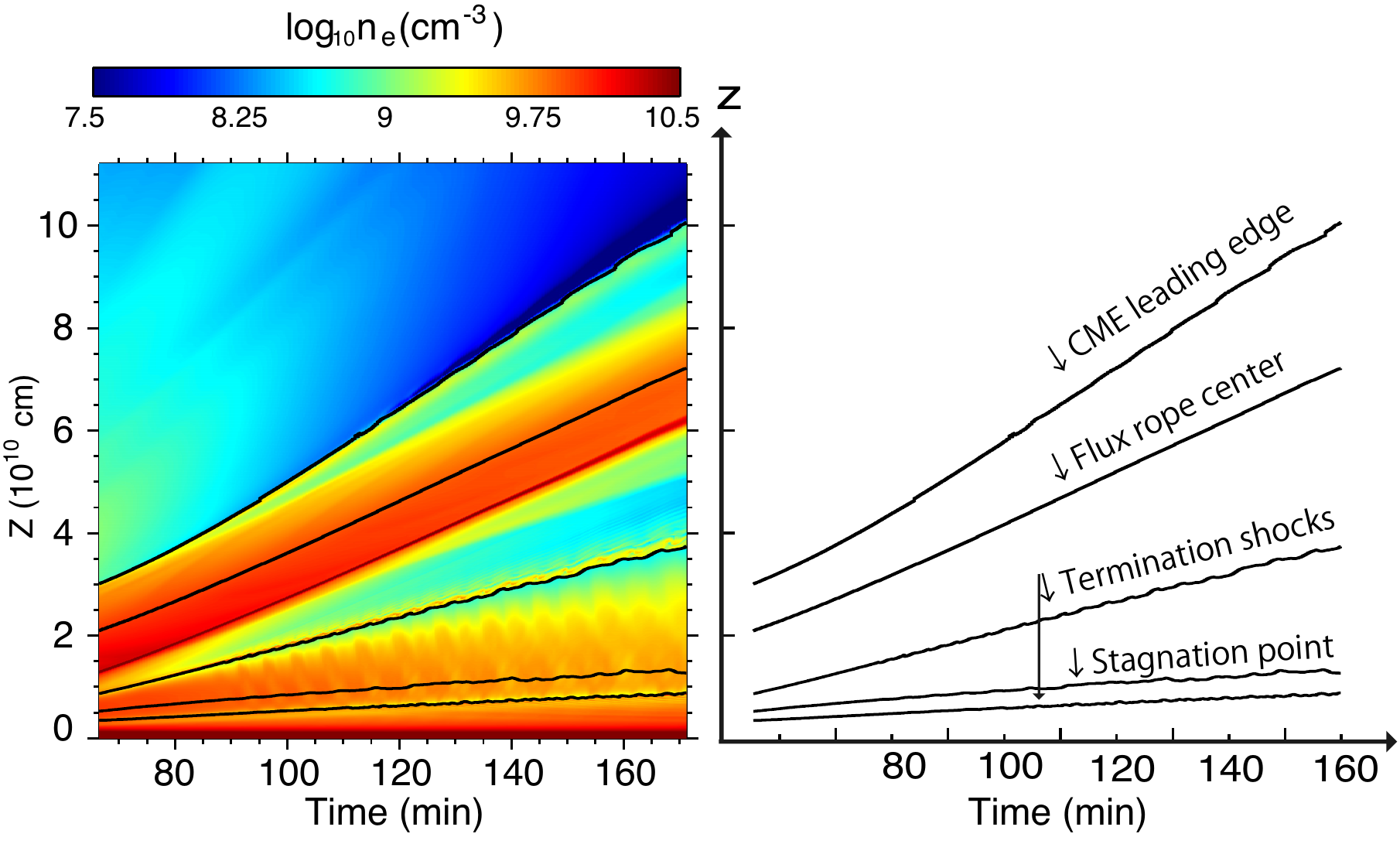}
\caption{The result of automatic detection of the height of the CME leading edge, flux rope center, termination shocks and the stagnation point after the time $t=65$ minutes along $x=0$ line in Run B. In the left panel, the automatically detected positions are plotted as black solid lines over the color-shaded time-distance plot of the electron number density in log scale. In the right panel, the height-time profiles of the detected structures are shown.}
\label{flare}
\end{figure}

The right Figure 11 shows the automatic measurements after the time $t=65$ minutes. In the left panel of Figure 11, these measurements are
plotted over the time-distance plot of the electron number density along $x=0$ line. The height of the CME leading edge is 
detected as the sharp change in the electron number density. The center of the flux rope is detected as the location at which 
the magnetic flux function takes its maximum value. The termination shocks at both ends of the current sheet are detected as 
the discontinuity of $v_z$. The stagnation point is detected as the place where $v_z$ vanishes along $x=0$. By those methods, 
we are able to track automatically the evolution of these structures after time $t=65$ minutes. These measurements show clearly
the oscillations of the height of the CME leading edge and the termination shocks seen in Figure 10.

\begin{figure}
\includegraphics[width=1.0\textwidth]{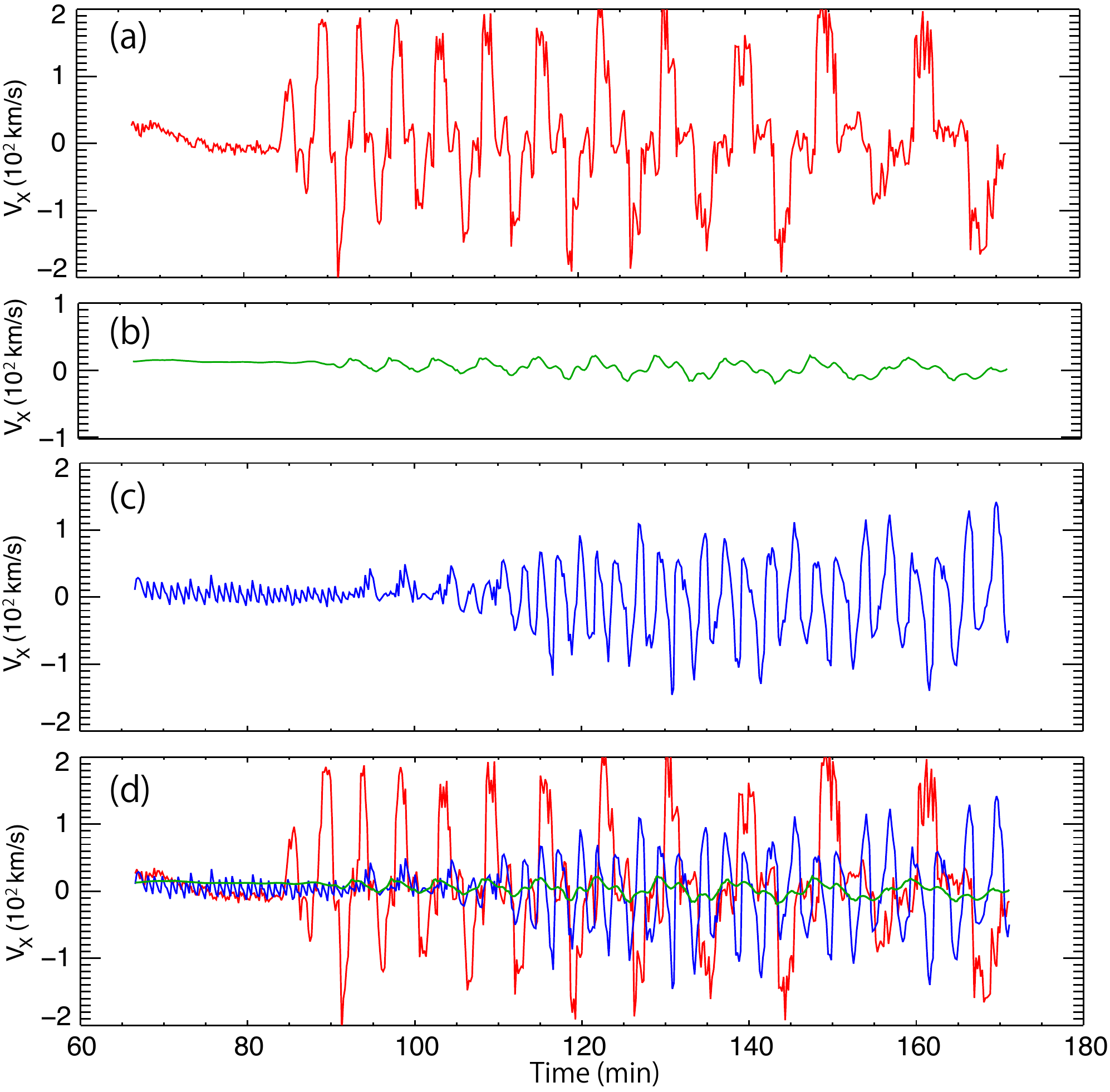}
\caption{The time variation of $v_x$ at the buffer region (a), the stagnation point (b), and the loop top (c). All the time plots are overplotted in panel (d).}
\label{flare}
\end{figure}

The red solid line in Figure 12 (a) shows the time variation of $v_x$ at the buffer region (just above $z_{TS+}$) after $t=65$ minutes. We see 
that the buffer oscillation starts abruptly at time $t=83$ minutes. The apparent period of the buffer oscillation at 
the beginning is $\sim5$ minutes, and it grows to be around $\sim12$ minutes at $t\simeq 160$ minutes. 
The time variation of $v_x$ at the flare loop top, on the other hand, is shown in the blue solid line in Figure 12 (c). 
At around $t=94$ minutes, a quasi-periodic oscillation begins with a period of $\sim 5$ minutes, and at $t\sim 110$ minutes, 
the period of the oscillation changes to $\sim 2$ minutes. The $\sim 5$-minutes loop top oscillation is caused by the horizontal 
wave that propagates from the buffer region at the bottom of the flux rope. This can be seen in the time-distance plot of $v_x$ 
in Figure 7. The subsequent oscillation of period of $\sim 2$minutes actually occurs locally at the loop top. Figure 12 (b) shows 
the time variation of $v_x$ at the stagnation point at $z=z_{SP}$. Figure 12 (d) shows $v_x$ at those three different places altogether. 

\begin{figure}
\includegraphics[width=1.0\textwidth]{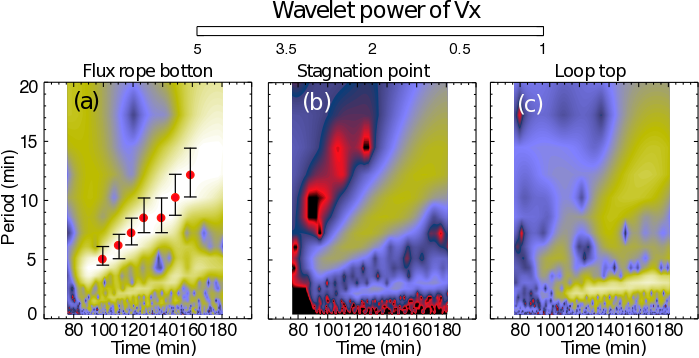}
\caption{The wavelet power of the time variation of $v_x$ at the buffer region (a), at the stagnation point (b), and at the flare loop top (c) color shaded in logarithmic scale. The period with peak oscillation power in the buffer regions are denoted by red filled circles in panel (a) at times $t=$100, 110, 120, 130, 140, 150 and 160 minutes, respectively. The error bars accompanying the red circles correspond to the resolution in frequency in wavelet transformation.}
\label{flare}
\end{figure}

We then apply a wavelet analysis to the time variation of $v_x$ at the bottom of the flux rope, the stagnation point, and the flare loop top. 
Figure 13 (a), (b) and (c) show the wavelet power of the time variation of $v_x$ at those three different heights. In Figure 13 (a), a 
strong enhancement of the wave power at the period of $P\simeq 5$ minutes is present at $t\simeq 90$ minutes after the start of the eruption. 
The oscillation period of the peak wavelet power increases with time, and becomes $P\simeq12$ minutes at time $t=160$ minutes. This is the 
result of the buffer oscillation at the bottom of the flux rope.
In Figure 13 (c), we see the oscillation at the flare loop top in the wavelet space. A clear enhancement of the wavelet power at 
the period of $\sim 2$ minutes appeared at the time $t\simeq115$ minutes. The period gradually increases with time and becomes $P\simeq3$ 
minutes at time $t=160$ minutes. In Figure 13 (c), we can also see an enhancement of the wavelet power at the flare loop top whose peak period 
increases from $P\simeq5$ minutes due to the arrival of the waves propagating from the oscillating buffer region beneath the flux rope. 
In Figure 13 (b), we show the wavelet power of $v_x$ at the stagnation point, which also exhibits two oscillation patterns starting from 5 and 2 minutes, respectively. 

The oscillation period $P_{buffer}$ is roughly determined by the time scale needed for the flow to go back and forth within the buffer 
region as $P\sim 2L_{buf}/v_{buf}$, with $L_{buf}$ and $v_{buf}$ being the size of the buffer region and the flow speed, respectively. 
From Figure 12, the amplitude of the horizontal speed of the buffer oscillation is $\sim 2\times10^2$ km s$^{-1}$.
The reconnection outflow that enters the buffer region is once thermalized by horizontal and oblique shocks and 
accelerated again by the pressure gradient force. This results in the velocity amplitude to be of order of the sound speed 
at the buffer region, which is of the same order of the Alfven speed in the inflow region at the height of the stagnation point 
$v_{buf}\sim C_{A,in}$, with $C_{A,in}$ being the Alfven speed in the inflow region. We note that the characteristics of oscillating supersonic back flows is quite similar to the flare loop top oscillation discussed in \citet{takasao2016}.

If we assume that the length scale of the buffer region is proportional to the width of the reconnection outflow at the exit near the buffer region ($L_{buf}\sim w$), the period of the buffer oscillation is expected to be proportional to $w/C_{A,in}$. 
Then, we define the time average of a quantity $q$ at time $t$ with an averaging time window $\tau$ as follows,
\begin{equation}
\bar{q}_{\tau}(t)=\frac{1}{\tau}\int_{t-\tau/2}^{t+\tau/2} q(t')dt'.
\end{equation}

{\bf{Figure 14 shows the time evolution of the period at which the wavelet power of the buffer oscillation in $v_x$ have its peak.
We note that the buffer oscillation period is consistent with the relation $P\simeq17.5\bar{w}_{\tau}/\bar{C}_{A,\tau}$ indicated with black solid curve in Figure 14.}} In the above analysis, we chose $\tau=19$ minutes in the averaging procedure to remove the fluctuation whose time scale is shorter than the buffer oscillation period. The time resolution of the simulation data we analyzed is $0.01\tau$.

\begin{figure}
\includegraphics[width=1.0\textwidth]{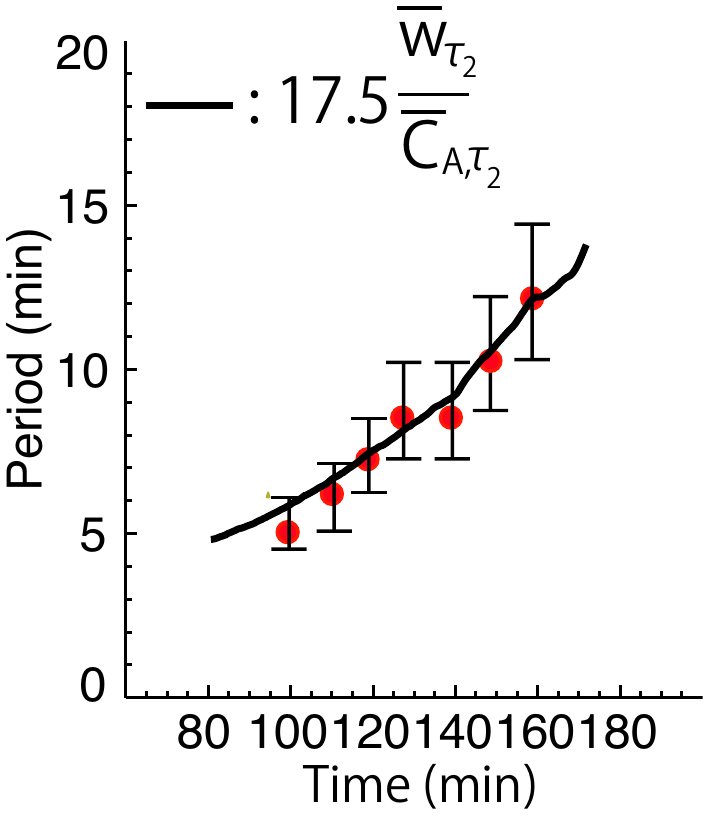}
\caption{The time evolution of the period at which the wavelet power of the buffer oscillation in $v_x$ have its peak (red filled circles) with error bars. The error bars correspond to the frequency resolution of the wavelet transformation. The phenomenological relation $P\simeq 17.5\bar{w_{\tau_2}}/\bar{C_{A,\tau_2}}$ is over-plotted with a thick solid line.}
\label{flare}
\end{figure}

\begin{figure}
\includegraphics[width=1.0\textwidth]{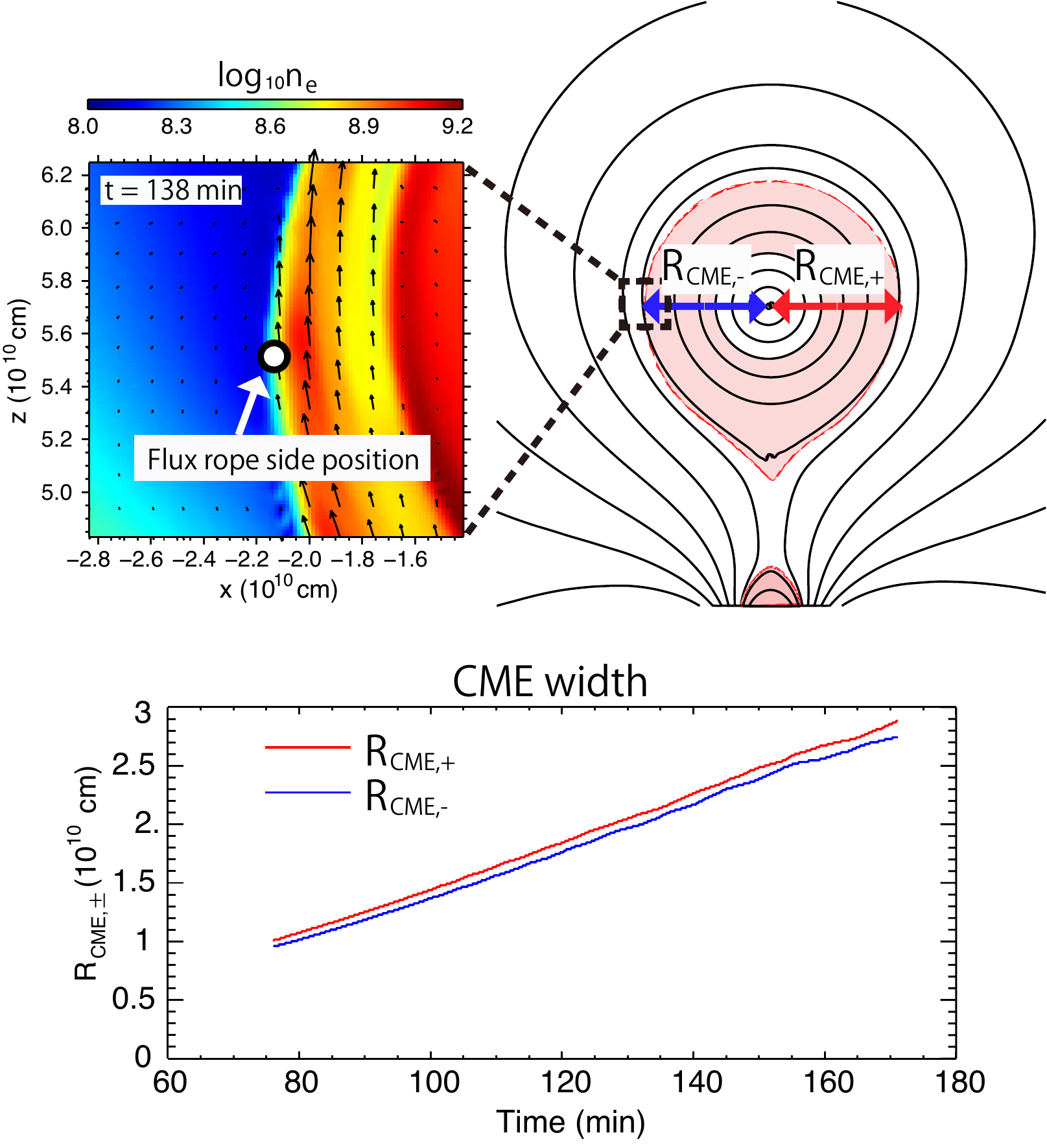}
\caption{The definition and the time evolution of the lateral CME edge position. Panel (a) shows the electron number density and plasma flow field at the flux rope edge. Panel (b) schematically shows the definition of the CME edge positions. Red and blue solid lines in panel (c) show the time evolution of the distances of automatically detected lateral edges flux rope from $x=0$ line.}
\label{flare}
\end{figure}

\section{The lateral oscillation of the CME flux rope}
In order to discuss the impact of the buffer oscillation on the oscillation of the global CME flux rope, we 
discuss the time evolution of the lateral extent and the leading edge position of the CME. First, we define the 
lateral extent of the CME flux rope as the boundary between the background corona and the 
upward-propagating plasma surrounding the CME ejecta at the height of the flux rope center.
We indicate the position (x-coordinate) of the boundaries on the positive and negative sides of the CME as $R_{CME,+}$ and $-R_{CME,-}$, respectively. We automatically detect the boundaries as a sharp gradient of both $v_z$ and mass density $\rho$ in the horizontal direction (the reconnection jet has a higher $\rho$ and larger $v_z$). 
The upper panel of Figure 15 schematically shows the boundary position.
The time evolution of the lateral extent of the CME in the positive and negative sides $R_{CME,\pm}$ is shown in the bottom panel of Figure 15.

\begin{figure}
\includegraphics[width=1.0\textwidth]{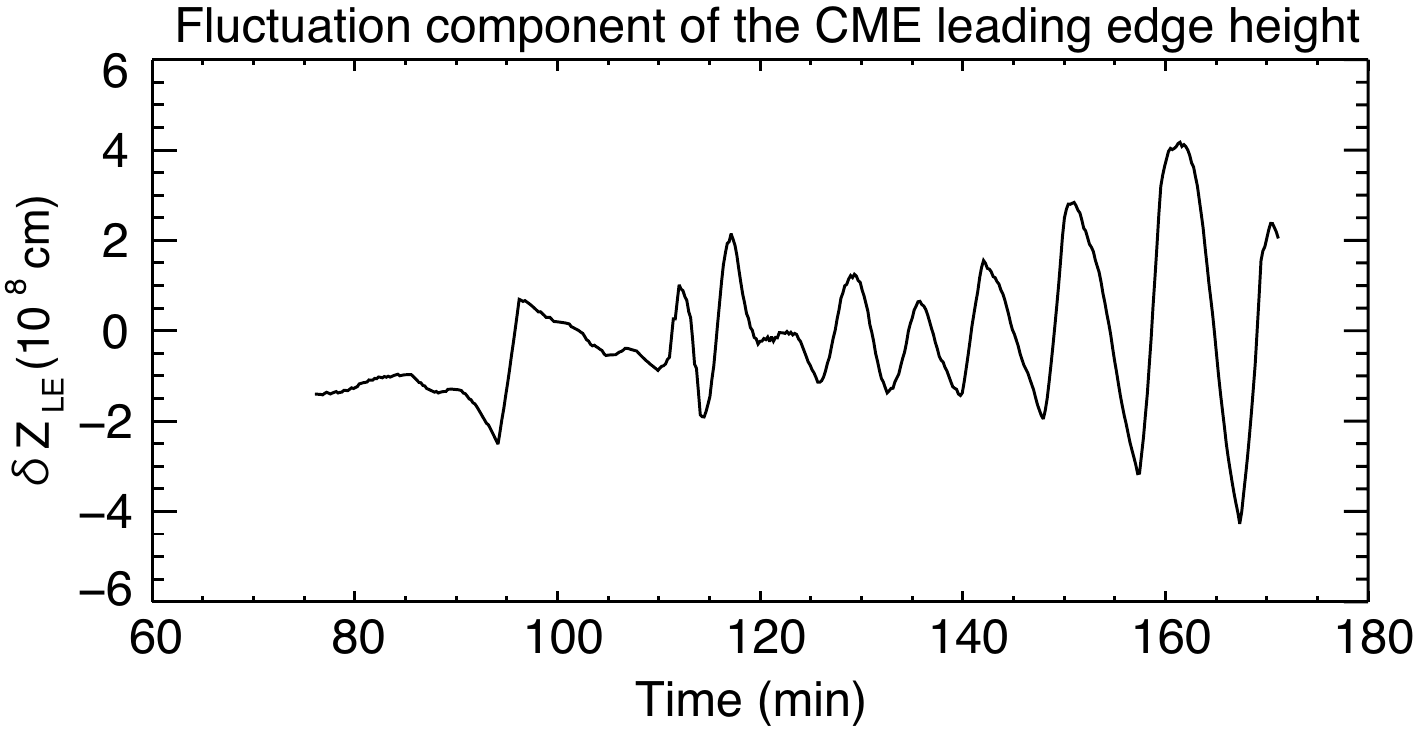}
\caption{The time evolution of the fluctuating component of the CME leading edge height between $t=75$ minutes and $t=172$ minutes.}
\label{flare}
\end{figure}

\begin{figure}
\includegraphics[width=1.0\textwidth]{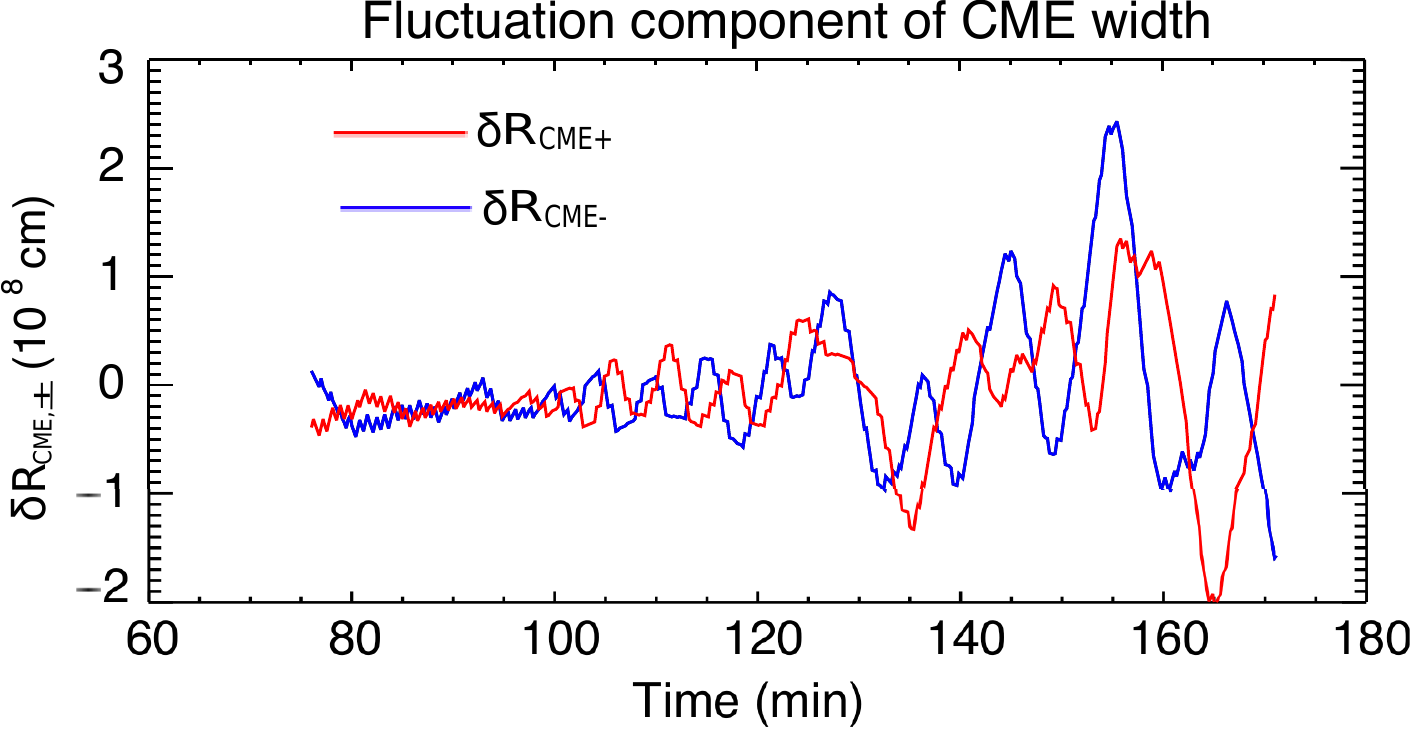}
\caption{The time evolution of the fluctuating component of the lateral extent of CME at both sides.}
\label{flare}
\end{figure}

\begin{figure}
\includegraphics[width=1.0\textwidth]{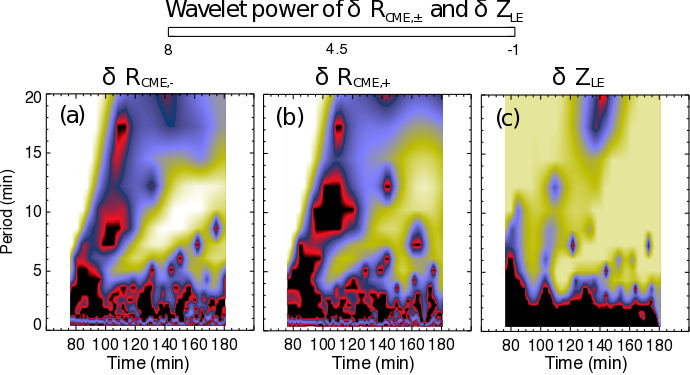}
\caption{The wavelet power of the fluctuation component of the lateral extent (panels (a) and (b)) and the leading edge position (panel (c)) of the CME.}
\label{flare}
\end{figure}

In order to see the oscillation of the CME flux rope surface, we define the fluctuating component of the position X at time $t$ as follows,
\begin{equation}
\delta X(\tau_1,\tau_2;t) =\bar{X}_{\tau_1}(t)-\bar{X}_{\tau_2}(t),
\end{equation}
where $\bar{X}_{\tau}(t)$ denotes the time average of the quantity $X$ at time $t$ with an averaging time window $\tau$ defined in the previous section.

$\delta X(\tau_1, \tau_2; t)$ takes out the variation of $X$ with the time scale between $\tau_1$ and $\tau_2$. 

Figure 16 shows the time evolution of $\delta z_{LE}(\tau_1,\tau_2;t)$ with averaging time windows $\tau_1=0.1\tau=1.9$ minute and $\tau_2=\tau=19$ minutes. 
A clear oscillation signature of the leading edge position appears at time $t\sim 120$ minutes. The oscillation period increases from $P\sim7$ minutes to $\sim 12$ minutes. The oscillation amplitude increases with time from $\Delta\sim 2\times10^8$ cm to $4\times10^8$ cm. The corresponding velocity amplitude of the oscillation in $\delta z_{LE}$ is $\sim \Delta/P\sim5\times10^5$ cm s$^{-1}$.

The lateral oscillation of the CME is evident in the fluctuation of $R_{CME,\pm}$. In Figure 17, we plot $\delta R_{CME,+}(\tau_1,\tau_2;t)$ 
and $\delta R_{CME,-}(\tau_1,\tau_2;t)$ with averaging windows $\tau_1=0.1\tau=1.9$ minutes and $\tau_2=\tau=19$ minutes in red and blue solid lines, respectively. At $t\sim100$ minutes, an oscillation with a period of $P\sim 7$minutes sets off, and the oscillation period increases with time to be $P\sim 12$ minutes at time $t\simeq 160$ minutes. The oscillations in $R_{CME,+}$ and $R_{CME,-}$ are observed to be out of phase all the time. 
The oscillation amplitude also increases with time from $\Delta\sim 5\times10^7$ cm to $2\times10^8$ cm. The corresponding velocity amplitude of the oscillation in $\delta R_{CME,\pm}$ is $\sim \Delta/P\sim2\times10^5$ cm s$^{-1}$.
Figure 18 (a), (b) and (c) show the wavelet power of $R_{CME,-}$, $R_{CME,+}$ and $z_{LE}$, respectively. Strong enhancement of the wavelet power is present at $t\simeq 100$minutes at the period of $P\sim 5$minutes and the peak period grows with time to reach $P\sim 12$minutes at 
time $t\simeq 120$ minutes. In Figure 18 (c), on the other hand, the enhancement of wave power appears at time $t\sim 115$ minutes at 
the period of $P\sim 5$minutes, and the peak period increases with time to reach $P\sim 12$minutes at time $t\simeq160$ minutes. 

\section{Comparison with quasi-periodic pulsations in an eruptive flare}
\citet{brannon2015} reported sawtooth-like substructures within flare ribbons that showed quasi-periodic oscillations in their positions in an eruptive M-class flare observed by IRIS on 2014 April 18. The sawtooth pattern seen in the IRIS Si {\sc iv} passband 
maintained its shape for twenty minutes, drifting back and forth along the flare ribbon with a speed of $\sim15$ km s$^{-1}$ in 
the plane of sky and a period of $\sim 140$ s, respectively. Doppler shifts are also measured in the Si {\sc iv} line at the 
location of the ribbon oscillation, showing the oscillation also in the line-of-sight
velocity with an amplitude of 20~km s$^{-1}$. \citet{brannon2015} discussed the tearing mode or Kelvin-Helmholtz instability in the flare current sheet to be a possible mechanism for driving the sawtooth oscillation pattern. \citet{brosius2015, brosius2016} have also reported oscillations 
in the light curves and Doppler-shifts of other chromospheric lines observed by IRIS and the {\em Extreme ultraviolet 
Imaging Spectrometer (EIS) onboard Hinode, and in the HXR light curves observed
by the {\em Ramaty High Energy Solar Spectroscopic Imager (RHESSI).

The flare is also observed by the {\em Atmospheric Imaging Assembly} (AIA) on the {\em Solar Dynamic Observatory} (SDO). 
It is accompanied by a fast CME observed by the {\em Solar Terrestrial Relations Observatory} (STEREO) and 
the {\em Large Angle and Spectrometric Coronagraph} (LASCO) on {\em Solar and Heliospheric Observatory} (SOHO) 
with a much lower tempo-spatial resolution. In this section, we report the quasi-periodic pulsation shown 
in the total light curves of the flare emission in the soft X-ray (SXR) and HXR in the context of magnetic
reconnection and CME evolution, and compare these QPPs with the simulation results.
Figure 19 (a) shows the flare ribbons plotted over the line-of-sight magnetogram taken by the {\em
Helioseismic and Magnetic Imager} (HMI) on SDO. The pixels that brightened in the SDO/AIA 1600\AA\ passband 
are taken as the feet of reconnection-formed flare loops and color-coded with the lapsed time. At a given time during
the flare, the reconnected magnetic flux is measured by summing up magnetic flux encompassed by the brightened flare
ribbons, and its time derivative gives the rate of magnetic reconnection. These are plotted as red and blue 
lines respectively in Figure 19 (b), and uncertainties are derived from the difference between the reconnection flux
measured in the positive and negative magnetic fields. 
The lift-off of the CME is seen clearly in the
time-distance plot constructed from coronagraph images taken by COR1 of STEREO-A along the CME flight path 
(Figure 19 (c)), and the CME height ($H_{CME}$) projected to the sky plane is measured from the time-distance diagram 
and shown in Figure 19 (d). Figure 19 (d) also shows the light curves of the SXR emission at 1-8 ~\AA\ obtained by 
GOES ($F_{SXR}$) with a time cadence of 3~s and the HXR counts rate at photon energy 25 - 50~keV by RHESSI ($F_{HXR}$) 
with a cadence of 4~s. All these light curves show the flux integrated through the entire active region. Also plotted 
is the time derivative of $F_{SXR}$ ($\dot{F}_{SXR}$). The soft X-ray is emitted by hot flare plasmas in coronal 
loops at the temperature of a few to a few tens MK, and the 25 - 50~keV HXR photons are emitted from the foot-points 
of flare loops (see \citet{brosius2016} for HXR images).

Seen from the figure, quasi-periodic pulsations in $F_{HXR}$ is clearly recognized. Remarkably, the time 
derivative of $F_{SXR}$ ($\dot{F}_{SXR}$) also shows a clear oscillatory pattern starting from around 12:50, 
which continues for half an hour. During this time, the magnetic reconnection rate has reached the maximum
of 10$^{19}$ Mx s$^{-1}$, and the CME is rapidly accelerated to nearly 1000 km s$^{-1}$. 
The periods of the oscillations in these light curves are comparable with 
each other, and are also comparable with those reported by \citet{brannon2015} and \citet{brosius2015, brosius2016}, 
although observations are obtained by many different instruments. Oscillations shown in the time derivative 
of the GOES soft X-ray light curve are nearly in phase with the oscillations in the HXR light curve in the rise 
of the SXR emission, which demonstrates the Neupert effect in this flare \citep{neupert1968,dennis2003}. 
These observations suggest that, in this eruptive flare, QPPs are global signatures. 

To extract the observed oscillation signatures and compare with the numerical simulations, we also plot $\delta \dot{F}_{SXR}$, which is the smoothed $\dot{F}_{SXR}$ following the definition of Equation (26) with $\tau_1=4$ s and $\tau_2=4\times10^2$ s.  
Figure 20 shows the time variation of the wavelet power of $\delta{\dot{F}_{SXR}}$ after 12:37UT. The quasi-periodic variation seen in Figure 19 is illustrated by enhanced wave power in the period range between 0.5 minutes and 2 minutes. We can see two linear trails of the enhanced wave power that starts at 12:47~UT and lasts for about 25 minutes with a gradually growing period. The behavior is similar to the behavior of the loop top oscillation in the wavelet space as shown in Figure 13 (c).

\begin{figure}
\includegraphics[width=1.0\textwidth]{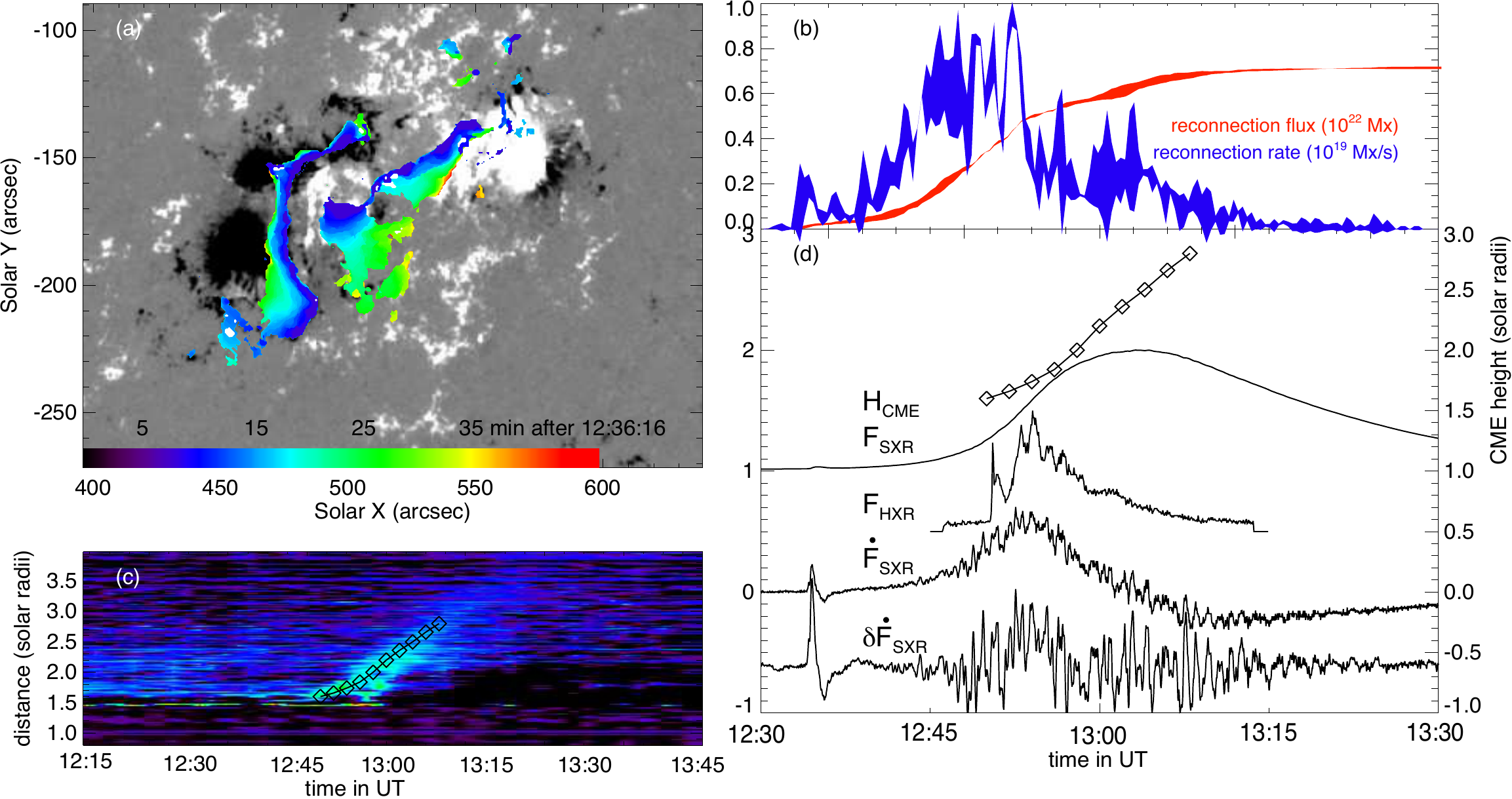}
\caption{(a) The flare ribbons superimposed on a line-of-sight magnetogram taken by SDO/HMI. The pixels that brightened in SDO/AIA 1600\AA passband are plotted, and the color code indicates the time (minutes after 12:36~UT) when the pixel is first brightened. (b) The time evolution of 
reconnected magnetic flux (red curve) and the rate of magnetic reconnection (blue curve) in the units of Mx and Mx $s^{-1}$, respectively. 
Reconnected magnetic flux at a given time is measured by summing up the magnetic flux in the area that have brightened in AIA 1600\AA 
images as shown in panel (a). (c) The time-distance diagram constructed along the flight path of the CME
from the base-difference images taken by the coronagraph COR1 on STEREO-A. The horizontal axis is the time and the vertical axis is 
the heliocentric location from the solar center in the units of solar radii. The bright structure that appeared at around 12.8~UT 
and moved upward in the plot shows the propagation of the CME leading edge. The time evolution of the CME height ($H_{CME}$) is shown as diamonds.
(d) Light curves of soft X-ray 1-8\AA\ pass band ($F_{SXR}$) by GOES, hard X-ray 25 - 50 keV energy band ($F_{HXR}$), 
and the time derivative of $F_{SXR}$ ($\dot{F}_{SXR}$) after 12:30~UT of April 18, 2014. $F_{HXR}$ is plotted after 12:45~UT. 
$\delta \dot{F}_{SXR}(\tau_1,\tau_2;t)$ with $\tau_1=4$ s and $\tau_2=4\times10^2$ s is also plotted at the bottom in order to emphasize 
the quasi-periodic variation in $\dot{F}_{SXR}$. All these light curves are arbitrarily normalized. 
The time evolution of the CME height ($H_{CME}$) is also shown in the units of solar radii (right axis).}
\label{flare}
\end{figure}

\begin{figure}
\includegraphics[width=0.7\textwidth]{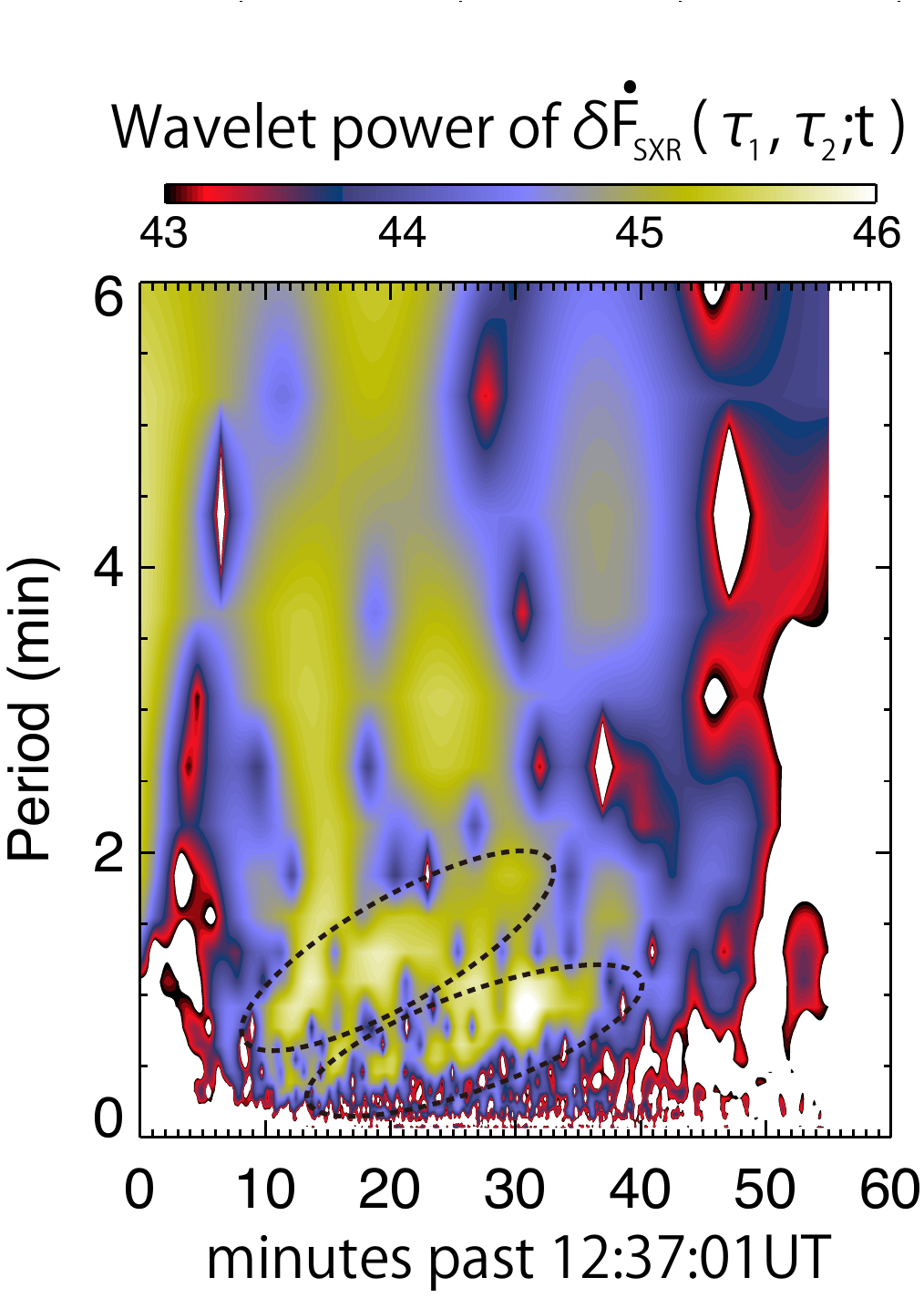}\centering
\caption{Wavelet power of $\delta \dot{F}_{SXR}(\tau_1,\tau_2;t)$ of the flare on 2014 April 18 with $\tau_1=4$ s and $\tau_2=4\times10^2$ s. Two linear trails of enhanced powers are indicated by dashed circles.}
\label{flare}
\end{figure}

\section{Summary and discussion}
We study dynamics of magnetic reconnection in eruptive solar flares using 2D numerical MHD simulations. Three simulation
runs (Runs A, B, and C) are conducted with three different Lundquist numbers, $S = $ $2.8\times10^3$, $5.6\times10^3$ and $2.8\times10^4$, respectively. 
In Run A (low Lundquist number case), the eruption process is laminar and no oscillation is generated throughout the simulation. 
In Run B, the collision of magnetic reconnection jets excites quasi-periodic oscillations at both ends of the current sheet whose 
periods grow longer with time. In Run C, quasi-periodic oscillations similar to the ones in Run B are excited, while the reconnection 
dynamics become more turbulent due to the progress of plasmoid instability in the current sheet. We call the oscillation driven by the 
collision of the reconnection outflow at the CME base the ``buffer oscillation''. The buffer oscillation period increases with time 
due to the expansion of the CME flux rope. The oscillation period of the CME surface is consistent with the period of the buffer oscillation. 

Oscillations of the CME have been observed, and the reported timescale ranges from ten to several hundred minutes at 
the heliocentric distance of $\sim 10 R_{\odot}$. Our numerical simulation does not cover such a distant. On the other 
hand, the buffer oscillation period is consistent with the phenomenological expectation of $P\simeq c_0w/C_A$, where 
$w$ is the width of the reconnection outflow near the CME flux rope, $C_A$ is the Alfven speed in 
the inflow region near the stagnation point, and $c_0$ is a non dimensional constant ($c_0\simeq17.5$ in this study). If we assume that the 
relation $P\simeq 17.5w/C_A$ holds even at the heliodistance of $\sim 10R_{\odot}$, the observed oscillation period gives the estimate 
of width of the reconnection outflow by $w\simeq 6\times10^{-2}C_AP$. If we assume $C_A\sim 10^3$ km s$^{-1}$ and $P\sim 10-100$ minutes, 
the estimated outflow width is $w\sim4\times10^4$km-$4\times10^5$ km, which is consistent with the reported
thickness of the observed current sheet trailing the CME \citep{lin2015}. In our simulation, the current sheet beneath the flux rope also oscillates 
with the period of a few to 12 minutes because of both the buffer oscillation and the flare loop top oscillations. The velocity amplitude of 
the current sheet oscillation is of order of 20 km s$^{-1}$. \citet{li2016} have recently observed the oscillation of the post-CME current sheet 
with a period of 11 minutes and an averege horizontal velocity of about 9.5 km s$^{-1}$. These measurements are similar to the results
from our simulations. 

We also compare the oscillations at the top of flare loops in our simulations with QPPs observed in the light curves of an 
eruptive flare on 2014 April 18 \citep{brannon2015, brosius2015, brosius2016}. QPPs are found in the HXR emission at the flare 
ribbons, as well as in the time derivative of the coronal SXR emission observed by GOES. 
The quasi-periodic signature of $\dot{F}_{SXR}$ is characterized by two distinct trails of enhanced wavelet power whose 
peak oscillation period grows with time. Such a pattern is very similar to the wavelet power-gram of the loop top 
oscillation generated in the simulation Run B. It is likely that the quasi-periodic oscillation of the termination shock 
at the top of the flare loop generates non-thermal electrons quasi-periodically, which then precipitate and heat the chromosphere, 
leading to QPPs in the observed HXR emission at flare ribbons, as well as the time derivative of the SXR emission (the Neupert effect). 
We note that the QPPs take place during the time when magnetic reconnection peaks and the CME rises rapidly. These results 
agree with the finding by \citet{kuznetsov2016} that HXR QPPs are often reported during eruptive events, suggesting
that the dynamical structure associated with eruptions (possibly, the formation of long thin current sheet) plays an important 
role in the generation of HXR QPPs.

The authors are grateful to Dr. Dana Longcope and Dr. Charles Kankelborg of Montana State University, Dr. Valery M. Nakariakov of University of Warwik, Dr. Shinsuke Takasao of Nagoya University and Dr. David McKenzie of NASA Marshall Space Flight Center for their fruitful comments and discussions which have made our original idea more concrete. The SDO/AIA data are courtesy of NASA/SDO and the AIA science team. The simulation code used in this work is created with the help of the HPCI Strategic Program. Numerical computations were carried out on Cray XC30 at the Center for Computational Astrophysics, National Astronomical Observatory of Japan. This work was financially supported by the Grant-in-Aid for JSPS Fellows 15J02548 and JSPS KAKENHI Grant Numbers 16H03955. This work was also supported by the “UCHUGAKU” project of the Unit of Synergetic Studies for Space, Kyoto University. 



\end{document}